\pdfoutput=1

\documentclass[11pt,twoside,a4paper,cmspaper,final,collab]{cms-tdr}
\def\svnVersion{282889}\def\cmsCernNo{2013-037}\def\cmsCernDate{\today}\def\cmsMessage{Published in Physical Review D as \href{http://dx.doi.org/10.1103/PhysRevD.91.052009}{\doi{10.1103/PhysRevD.91.052009.}}}

\begin{document}\cmsNoteHeader{EXO-12-059}

\hyphenation{had-ron-i-za-tion}
\hyphenation{cal-or-i-me-ter}
\hyphenation{de-vices}

\RCS$Revision: 274499 $
\RCS$HeadURL: svn+ssh://svn.cern.ch/reps/tdr2/papers/EXO-12-059/trunk/EXO-12-059.tex $
\RCS$Id: EXO-12-059.tex 274499 2015-01-21 10:59:23Z santanas $
\newlength\cmsFigWidth
\ifthenelse{\boolean{cms@external}}{\setlength\cmsFigWidth{0.98\columnwidth}}{\setlength\cmsFigWidth{0.48\textwidth}}
\ifthenelse{\boolean{cms@external}}{\providecommand{\cmsLeft}{top}}{\providecommand{\cmsLeft}{left}}
\ifthenelse{\boolean{cms@external}}{\providecommand{\cmsRight}{bottom}}{\providecommand{\cmsRight}{right}}
\providecommand{\PZpr}{\ensuremath{\cmsSymbolFace{Z}^\prime}\xspace}
\providecommand{\sba}{\ensuremath{\sigma\mathcal{B}A}\xspace}
\ifthenelse{\boolean{cms@external}}{\providecommand{\CL}{C.L.\xspace}}{\providecommand{\CL}{CL\xspace}}
\newlength\tabwid
\setlength\tabwid{3em}
\newcolumntype{C}{>{\centering\arraybackslash}m{\tabwid}}
\cmsNoteHeader{EXO-12-059} 
\title{Search for resonances and quantum black holes using dijet mass spectra in proton-proton collisions at \texorpdfstring{$\sqrt{s}=8$\TeV}{sqrt(s) = 8 TeV}}

\newcommand{\mjj}{\ensuremath{m_{\mathrm{jj}}}\xspace}
\newcommand{\mjjFAT}{\ensuremath{m^{\tiny Wide}_{\mathrm{jj}}}\xspace}
\newcommand{\dphijj}{\ensuremath{\abs{\Delta\phi{\mathrm{jj}}}}\xspace}
\newcommand{\detajj}{\ensuremath{\abs{\Delta\eta_{\mathrm{jj}}}}\xspace}
\newcommand{\dphijjFAT}{\ensuremath{\abs{\Delta\phi^\text{Wide}_{\mathrm{jj}}}}\xspace}
\newcommand{\detajjFAT}{\ensuremath{\abs{\Delta\eta^\text{Wide}_{\mathrm{jj}}}}\xspace}
\newcommand{\Qstar}{\ensuremath{\PQq^*}\xspace}
\newcommand{\Ustar}{\ensuremath{\PQu^*}\xspace}
\newcommand{\Dstar}{\ensuremath{\PQd^*}\xspace}
\newcommand{\Bstar}{\ensuremath{\PQb^*}\xspace}

\newcommand{\samplelumi}{19.7\fbinv\xspace}
\newcommand{\lumiUncert}{2.6\%\xspace}
\newcommand{\QCDMCscaleFactorDiWideJetMass}{1.23\xspace}
\newcommand{\chiSquare}{26.8\xspace}
\newcommand{\NDF}{35\xspace}
\newcommand{\jecUncert}{1\%\xspace}
\newcommand{\minMjjCut }{890\xspace}
\newcommand{\highestMass}{5.15\TeV\xspace}

\date{\today}

\abstract{A search for resonances and quantum black holes is performed
  using the dijet mass spectra measured in proton-proton collisions at
  $\sqrt{s}=8$\TeV with the CMS detector at the LHC. The data set
  corresponds to an integrated luminosity of 19.7\fbinv.  In a search
  for narrow resonances that couple to quark-quark, quark-gluon, or
  gluon-gluon pairs,
  model-independent upper limits, at 95\% confidence
  level, are obtained on the production cross section of resonances,
  with masses above 1.2\TeV. When interpreted in the context of specific
  models the limits exclude: string resonances with masses below
  5.0\TeV; excited quarks below 3.5\TeV; scalar diquarks
  below 4.7\TeV; \PWpr{} bosons below 1.9\TeV or
  between 2.0 and 2.2\TeV; \PZpr bosons below 1.7\TeV; and
  Randall--Sundrum gravitons below 1.6\TeV. A separate
  search is conducted for narrow resonances that decay to final states
  including \PQb quarks. The first exclusion limit is set
  for excited \PQb quarks, with a lower mass limit between 1.2 and 1.6\TeV depending on their decay properties.
  Searches are also carried out for wide resonances, assuming for the
  first time width-to-mass ratios up to 30\%, and
  for quantum black holes with a range of model parameters.
  The wide resonance search excludes axigluons and
  colorons with mass below 3.6\TeV, and color-octet scalars with mass
  below 2.5\TeV. Lower bounds between 5.0 and 6.3\TeV are set on the
  masses of quantum black holes.}

\hypersetup{%
pdfauthor={CMS Collaboration},%
pdftitle={Search for resonances and quantum black holes using dijet mass spectra in proton-proton collisions at sqrt(s)=8 TeV},%
pdfsubject={CMS},%
pdfkeywords={CMS, physics, search, exotica, dijet, resonance}}

\maketitle 

\section{Introduction}

We report on a search for new states decaying to dijets in
proton-proton (\Pp\Pp) collisions at a center-of-mass energy of
$\sqrt{s}=8$\TeV.  The data sample corresponds to an integrated
luminosity of \samplelumi collected with the CMS detector at the CERN LHC in
2012.  This analysis extends the search for new phenomena presented in
previous
CMS~\cite{Khachatryan:2010jd,Chatrchyan:2011ns,QBH2011,Chatrchyan:2013xva,CMS:2012yf,Chatrchyan:2013qhXX}
and ATLAS~\cite{ATLAS2010,Aad:2011aj,Aad:2011fq,ATLAS:2012pu,Aad:2014aqa}
publications. A review of experimental searches for new particles
in the dijet mass spectrum is presented in Ref.~\cite{Harris:2011bh}.

Many extensions of the standard model (SM) predict the existence of
new massive particles that couple to quarks or antiquarks (\PQq) and
gluons (\Pg). These new particles could produce resonant bumps in the dijet
invariant mass distribution associated with strong interaction
processes. A similar signature could be produced by quantum black
holes (QBH) that decay primarily to dijet final states.

Four studies are reported in this paper: (i) a search for narrow dijet
resonances using the inclusive mass spectrum, with different
sensitivities to the masses of $\PQq\PQq$, $\PQq\Pg$, and $\Pg\Pg$ final states; (ii) a dedicated
search for narrow resonances decaying to \PQb quarks; (iii) a search for
wide dijet resonances in the $\PQq\PQq$ and $\Pg\Pg$ final states; and (iv) a search for
QBHs decaying to two jets.

We interpret the results in the context of particles predicted by several representative models:
string resonances (S)~\cite{Anchordoqui:2008di,Cullen:2000ef}; scalar
diquarks (D)~\cite{ref_diquark}; excited quarks
(\Qstar)~\cite{ref_qstar,Baur:1989kv} including excited \PQb quarks
(\Bstar); axigluons (A)~\cite{ref_axi,Bagger:1987fz,Chivukula:2011ng}; color-octet
colorons (C)~\cite{ref_coloron}; the color-octet scalar (S8)
resonances~\cite{Han:2010rf}; new gauge bosons (\PWpr{} and
\PZpr)~\cite{ref_gauge} with SM-like couplings (SSM);
Randall--Sundrum (RS) gravitons
(\cPG)~\cite{RS,ref_rsg,Bijnens:2001gh}; and QBHs~\cite{MR,Calmet,qbh1}. More details on the
specific choices of couplings and the cross sections assumed
for the models considered can be found in Ref.~\cite{CMS:2012yf}.

Narrow resonances are considered to be those that have small natural widths compared to the
experimental dijet mass resolution.  We search for narrow $\PQq\PQq$ and $\Pg\Pg$ resonances using the predicted dijet resonance line shape of the RS
graviton model for the parameter choice
$k/\overline{M}_\text{Pl}=0.1$,
where $k$ is the unknown curvature
scale of the extra dimension and $\overline{M}_\text{Pl}$ is the
reduced Planck scale. This choice corresponds to a natural width
equal to 1.5\% of the resonance mass.

To search for narrow resonances decaying to \PQb quarks, the events are
divided into samples with zero, one or two jets identified as
originating from \PQb quarks. These samples are labeled 0\PQb, 1\PQb, and 2\PQb,
respectively. The sensitivity of the search to a given signal model depends on
whether the predicted 0\PQb and 1\PQb resonant samples are dominated by gluons or
quarks in the final state. Therefore two scenarios are considered:
resonances that decay predominantly into pairs of gluons or \PQb quarks
(``$\Pg\Pg$/$\PQb\PQb$") or resonances that decay predominantly into quark pairs only
(``$\PQq\PQq$/$\PQb\PQb$"). Dijet mass shapes appropriate to $\Pg\Pg$ resonances or $\PQq\PQq$ resonances are used in conjunction with $\PQb\PQb$ mass shapes.
The dijet mass shapes in each tag category are
weighted according to the expected gluon, quark, or \PQb-quark content.

Wide resonances are considered to be those where
the natural width is comparable to or larger than the experimental
dijet mass resolution.
The signature for a wide resonance would be a broad
enhancement in the dijet mass distribution.
Wide $\PQq\PQq$ and $\Pg\Pg$ resonances are considered using the dijet resonance line
shape of the RS graviton model with larger values of
$k/\overline{M}_\text{Pl}$, which correspond to natural widths up to
30\% of the resonance mass.

Using the same technique employed in the inclusive analysis,
we search for QBHs decaying to dijet
final states. The search is motivated by theories in which the
effective Planck scale in the
presence of extra dimensions ($M_\mathrm{D}$) is significantly smaller
than the nominal Planck
scale ($M_\text{Pl}\sim 10^{16}$\TeV), as for instance in the
Arkani-Hamed--Dimopoulos--Dvali (ADD) model \cite{add,add1} of flat
extra dimensions or the RS model~\cite{RS,ref_rsg,Bijnens:2001gh} of warped extra
dimensions. The dijet mass spectrum for QBHs is characterized by a peaking
structure, as a result of the opening of the QBH production threshold
for parton center-of-mass energies above the minimum mass
$M_\mathrm{QBH}^\text{min}$ of QBHs and the steeply falling parton luminosity at
higher center-of-mass energies. This shape differs from a resonance
line shape and is almost independent of both the number of extra
dimensions $n$ and the scale $M_\text{D}$.

\section{The CMS detector}
The central feature of the CMS apparatus~\cite{refCMS} is a
superconducting solenoid of 6\unit{m} internal diameter providing an axial
field of 3.8\unit{T}.  Within the field volume are located the silicon pixel
and strip tracker and the barrel and endcap calorimeters; a lead
tungstate crystal electromagnetic calorimeter and a brass and
scintillator hadron calorimeter.  An iron/quartz-fiber calorimeter
is located in the forward region, outside the field volume.  For
triggering purposes and to facilitate jet reconstruction, the
calorimeter cells are grouped into towers projecting radially outward
from the center of the detector. Events are filtered using a two-tier
trigger system: a hardware-based first level (L1) and a software-based
high-level trigger (HLT). The information from the individual detectors
is combined in a global view of the event, the particle-flow (PF)
event reconstruction~\cite{PFT-09-001-PAS,CMS:2010byl}, which attempts to identify
all the particles detected in a collision and to measure their momenta.
A more detailed description of the CMS detector, together with a
definition of the coordinate system used and the relevant kinematic
variables, can be found in Ref.~\cite{refCMS}.

\section{Event selection} \label{sec:EventSelection}

At least one reconstructed vertex is required within $\abs{z} <
24\unit{cm}$. The primary vertex is defined as the vertex with the
highest sum of squared transverse momenta for the associated tracks.

The PF algorithm is used to reconstruct the particles in the event and
to identify them as muons, electrons (with associated bremsstrahlung
photons), photons (unconverted and converted), or charged/neutral
hadrons. The PF candidates are clustered into jets using the
anti-\kt algorithm~\cite{antikt} with distance parameter of
0.5, implemented in the \textsc{FastJet} package~\cite{Cacciari:2005hq}. An
event-by-event jet-area-based
correction~\cite{jetarea_fastjet,jetarea_fastjet_pu,1748-0221-6-11-P11002}
is applied to remove the energy from additional collisions in
the same bunch crossing (pileup).
The jet momenta are further corrected using calibration constants
derived from simulations, test beam results, and pp collision
data~\cite{1748-0221-6-11-P11002}.  All jets in this analysis are
required to have transverse momentum (\pt) greater than 30\GeV and
absolute value of pseudorapidity ($\eta$) less than 2.5.  Jet identification criteria~\cite{JME-10-003-PAS} are
applied to the two jets in the event with the highest \pt (leading
jets), in order to remove spurious events associated with calorimeter noise.
The event is rejected if either of these two jets fails these
criteria.

Geometrically close jets are combined into ``wide
jets''~\cite{Chatrchyan:2011ns,CMS:2012yf}, which are used to measure
the dijet mass spectrum and search for new resonances and QBHs.  The wide jet
algorithm is inspired by studies using jet grooming
algorithms~\cite{Cacciari:2008gd,Krohn:2009th,Abdesselam:2010pt} and
is intended to reduce the sensitivity to gluon radiation from the
colored final state.  The two jets with largest \pt are used as
seeds. The Lorentz vectors of all other jets are then added to the
closest leading jet, if within $\Delta R=\sqrt{\smash[b]{(\Delta\eta)^2 +
  (\Delta\phi)^2}}<1.1$ (where $\phi$ is the azimuthal angle in
radians), to obtain two wide jets, which then compose the
dijet system. The background from $t$-channel multijet events is
suppressed by requiring the pseudorapidity separation of the two wide
jets (\detajj) to be less than 1.3. In addition, we require
that both wide jets are reconstructed in the region
$\abs{\eta}<2.5$. These requirements maximize the search sensitivity for
isotropic decays of dijet resonances in the presence of multijet
background~\cite{Khachatryan:2010jd}.

The L1 trigger used for this search requires that the scalar sum of the
jet \pt (\HT) be larger than 150\GeV.  Events satisfying
the L1 trigger are then filtered by the HLT which requires
that either of the two following trigger selections is satisfied:
the first trigger requires $\HT> 650$\GeV;
the second trigger requires that the invariant
mass of the dijet system (\mjj), computed using the same
algorithm employed at the reconstruction level, be greater than 750\GeV. In the second trigger, $\detajj<1.5$ is required.
Biases from the trigger requirements are avoided by requiring that the fully
reconstructed events have $\mjj>\minMjjCut\GeV$.
In this region, the combined efficiency of the L1 and HLT triggers is found to be more than 99.7\%.

To identify jets originating from the hadronization of \PQb quarks,
an algorithm is used that combines information on secondary
vertices and reconstructed lifetime. The set of conditions used
corresponds to the loose working point of the algorithm,
which is known as the combined secondary vertex (CSV) and described
in detail in Ref.~\cite{Chatrchyan:2012jua}. The performance is studied using samples of LHC data
enriched in or depleted of \PQb quarks, as well as simulated samples. The
algorithm is applied to the two leading jets, and events are categorized as
0\PQb, 1\PQb, or 2\PQb. This categorization allows differences in tagging efficiency between data
and simulation to be evaluated, as a function of jet \pt. Corrections
referred to as ``b-tagging scale factors'' are derived, which are applied to
the simulated samples used in the analysis to correct for the differences
observed between simulation and data.

\begin{figure*}[!htb]
  \centering
    \includegraphics[width=0.325\textwidth]{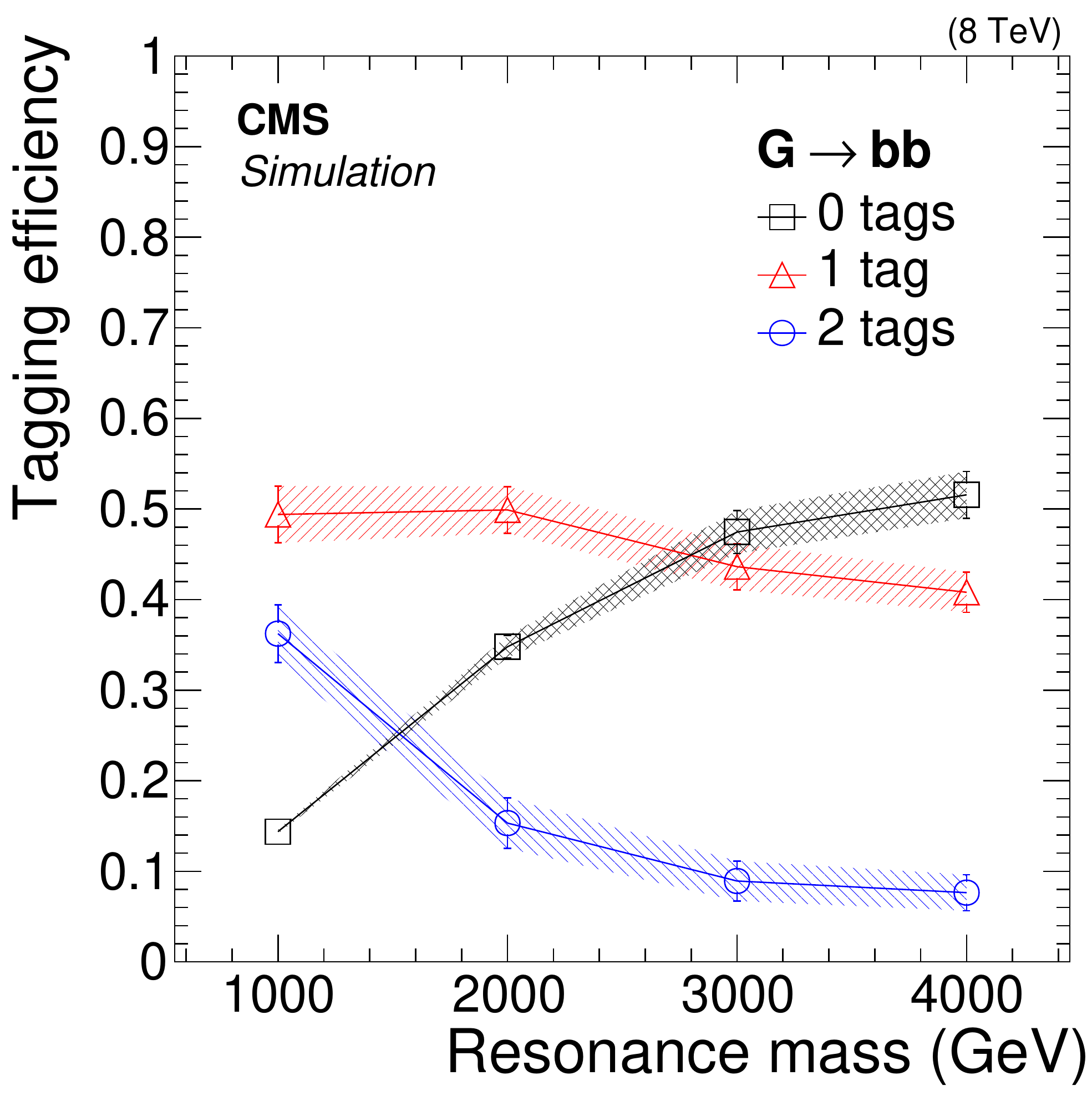}
    \includegraphics[width=0.325\textwidth]{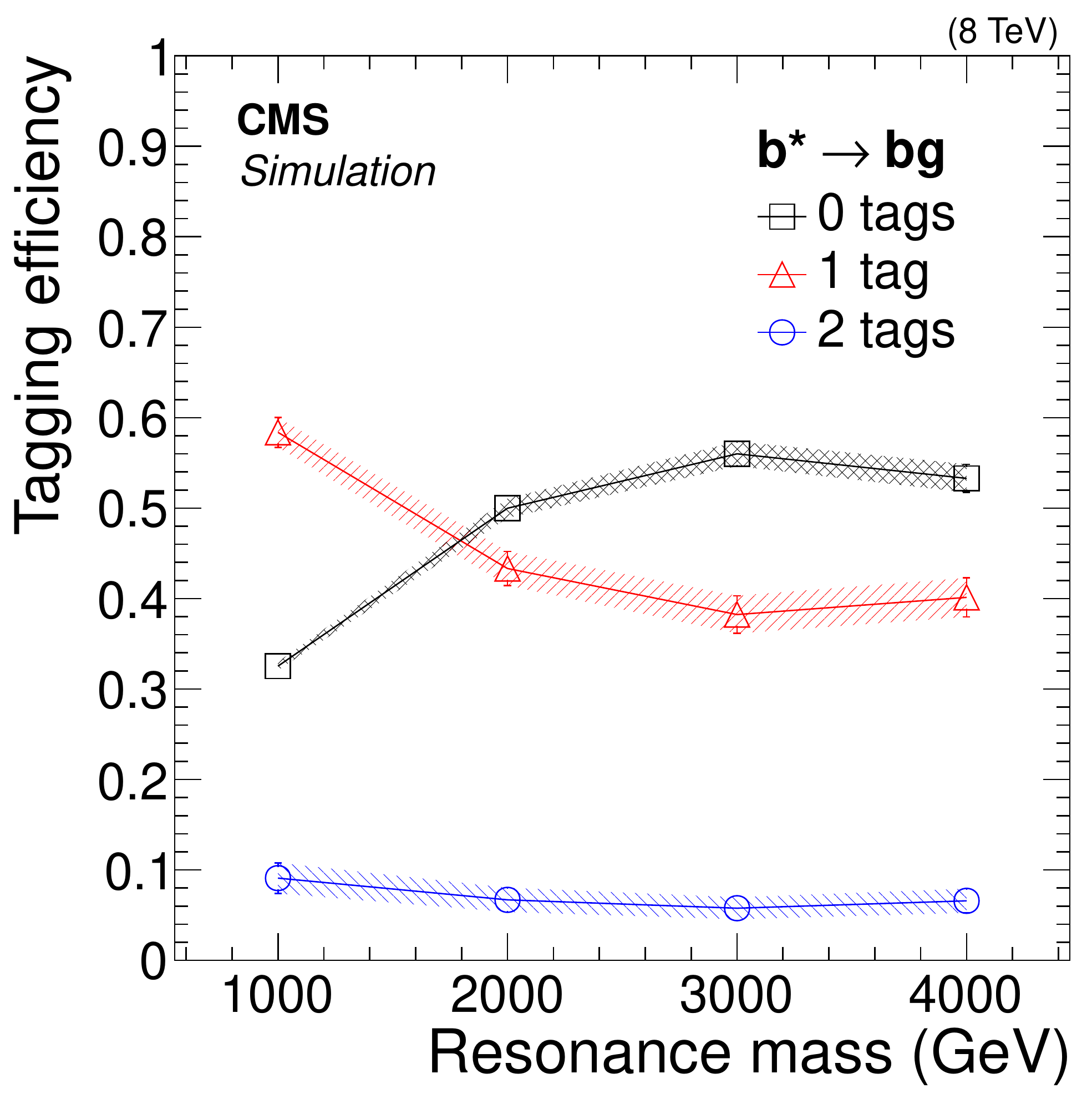}
    \includegraphics[width=0.325\textwidth]{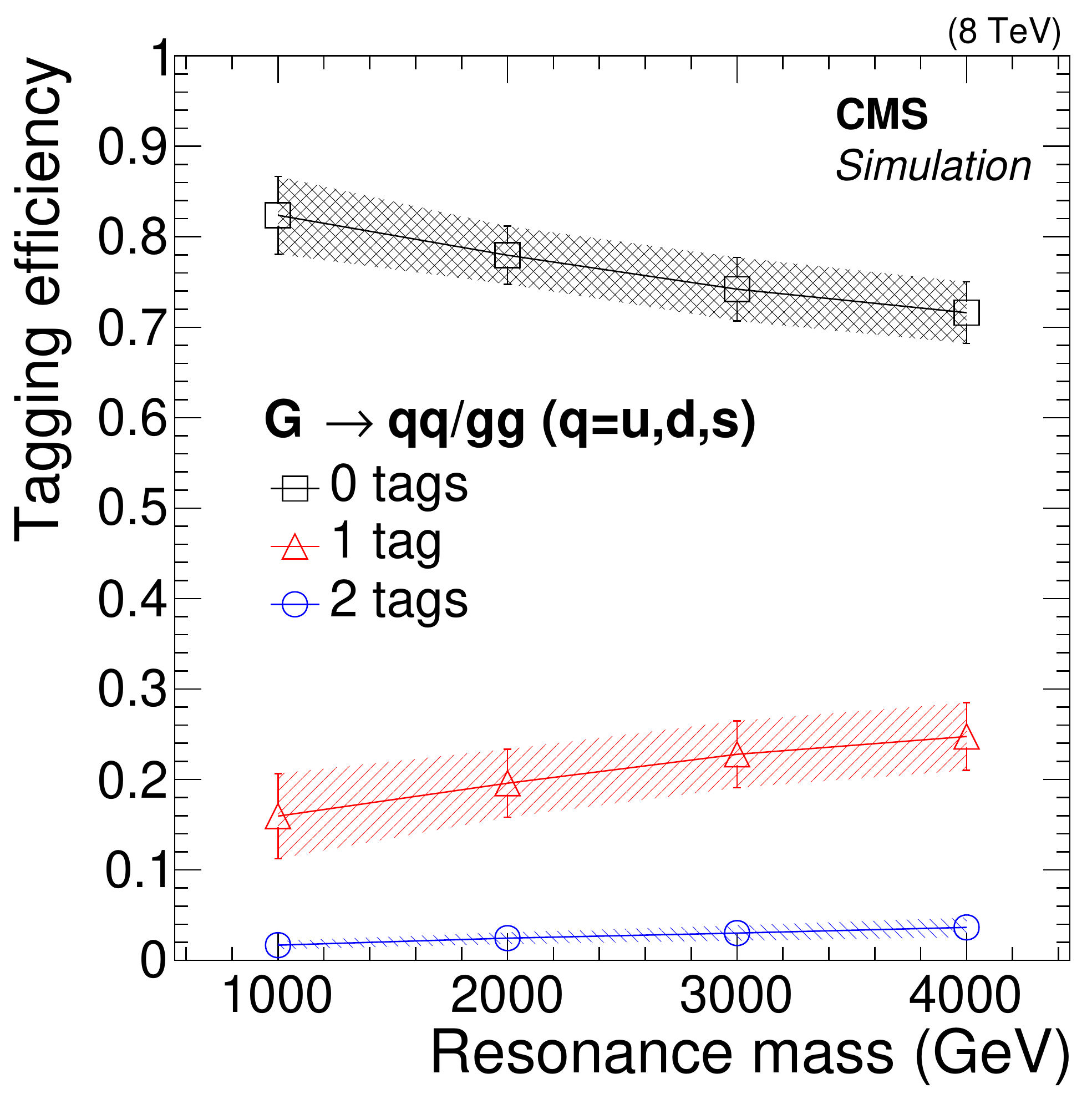}
    \caption{Tagging efficiencies for 0\PQb, 1\PQb, and 2\PQb selections as a
      function of the resonance mass for $\PQb\PQb$, $\PQb\Pg$, and
      $\PQq\PQq$/$\Pg\Pg$ (where $\PQq=\PQu$, \PQd, \PQs) decay modes, for an RS
      graviton \PXXG decaying to a $\PQb\PQb$ pair (left), an excited $\PQb^*$
      quark decaying to a \PQb-quark and a gluon (center) and an RS
      graviton \PXXG decaying to two gluons or to a $\PQq\PQq$ pair, with
      $\PQq=\PQu$, \PQd, or \PQs (right).  The hatched regions represent the
      uncertainties in the tagging efficiencies corresponding to the variation
      of the \PQb-tagging scale factors within their uncertainties.}
    \label{fig:rate}
  \end{figure*}

The tagging efficiencies for
0\PQb, 1\PQb, and 2\PQb categories are shown in Fig.~\ref{fig:rate} for RS
gravitons and excited \PQb quarks as a function of the resonance mass.
The efficiency to tag correctly a \PQb-jet decreases as the resonance
mass increases.  The efficiency of double-tagging a resonance that
decays into two charm quarks (${\sim}10\%$ at 1\TeV) is
systematically higher than that for $\Pg\Pg$, $\PQq\Pg$, and $\PQq\PQq$ final states with
light-flavor quarks (below ${\sim}5\%$), while being significantly
lower than for the $\PQb\PQb$ decay mode.  To simplify the analysis, charm
quarks are assumed to have the same efficiency as light quarks and
gluons. This choice translates into weaker sensitivity to $\PQb\PQb$ resonances.

\section{Dijet mass spectrum}

\begin{figure}[hbt]
  \centering
    \includegraphics[width=\cmsFigWidth]{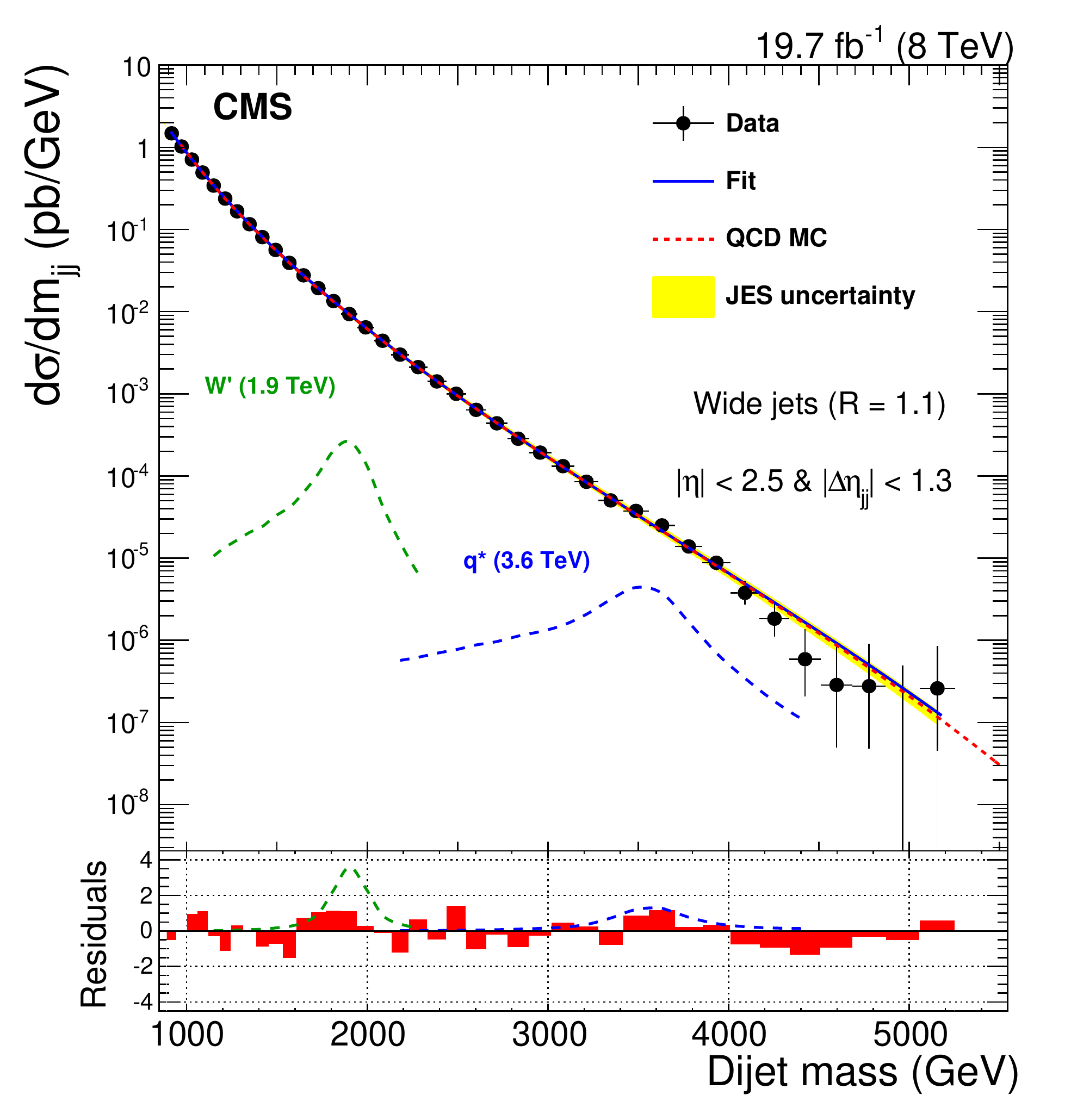}
    \caption{Inclusive dijet mass spectrum from wide jets
      (points) compared to a fit (solid curve) and to
      predictions including detector simulation of
      multijet events and signal resonances.  The predicted multijet
      shape (QCD MC) has been scaled to the data (see text).
      The vertical error bars are statistical only
      and the horizontal error bars are the bin widths.
      For comparison,the signal distributions for a \PWpr resonance of
      mass 1900\GeV and an excited quark of mass 3.6\TeV are shown.
      The bin-by-bin fit residuals scaled to the statistical uncertainty of the data,
      $(\text{data}-\text{fit})/\sigma_{\text{data}}$, are shown at the
      bottom and compared with the expected signal contributions.
    }
      \label{figDataAndMC}
  \end{figure}

Figure~\ref{figDataAndMC} shows the dijet mass distribution
normalized to the integrated luminosity of the sample
(${\rd\sigma}/{\rd \mjj}$) for the
inclusive data sample, with bin widths approximately equal to the dijet
mass resolution.
The data are compared to a leading order (LO) prediction of the
multijet background from \PYTHIA~{6.426}
\cite{refPYTHIA} with the Z2* tune~\cite{bib_pythiatunes} (Z2 is identical to Z1 aside from the choice of the CTEQ6L PDF),
where the generated events are processed through a \GEANTfour-based
\cite{refGEANT} simulation of the CMS detector.

In the event generation, CTEQ6L1 parton distribution functions
(PDF)~\cite{Pumplin:2002vw} are used.  The renormalization and
factorization scales are both set equal to the \pt of the
hard-scattered partons. The prediction has been normalized to the data
by applying a multiplicative factor of~\QCDMCscaleFactorDiWideJetMass.
The shape of the \PYTHIA prediction agrees with the data within the
statistical precision.

\begin{figure*}[!htb]
  \centering
    \includegraphics[width=0.48\textwidth]{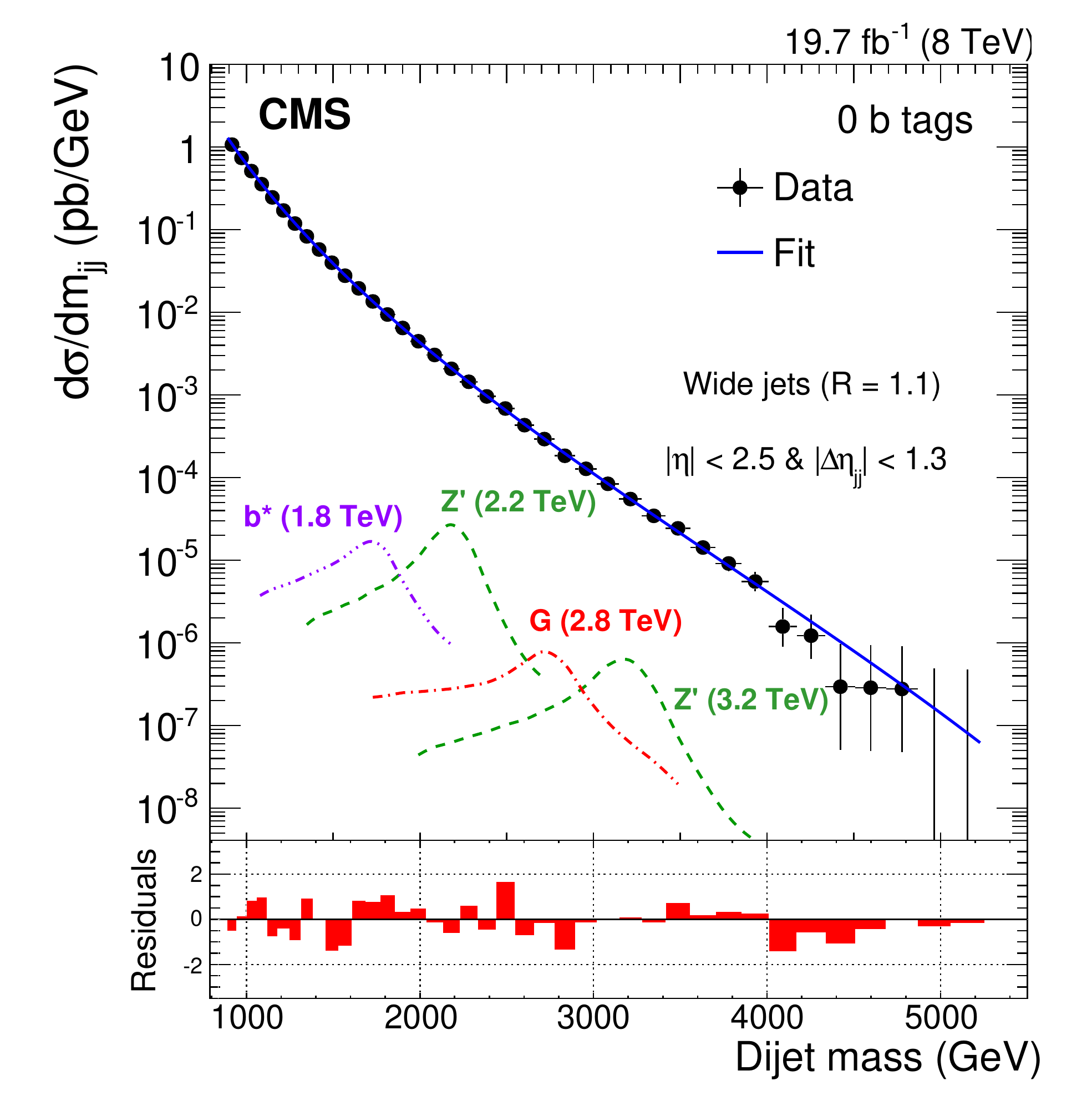}
    \includegraphics[width=0.48\textwidth]{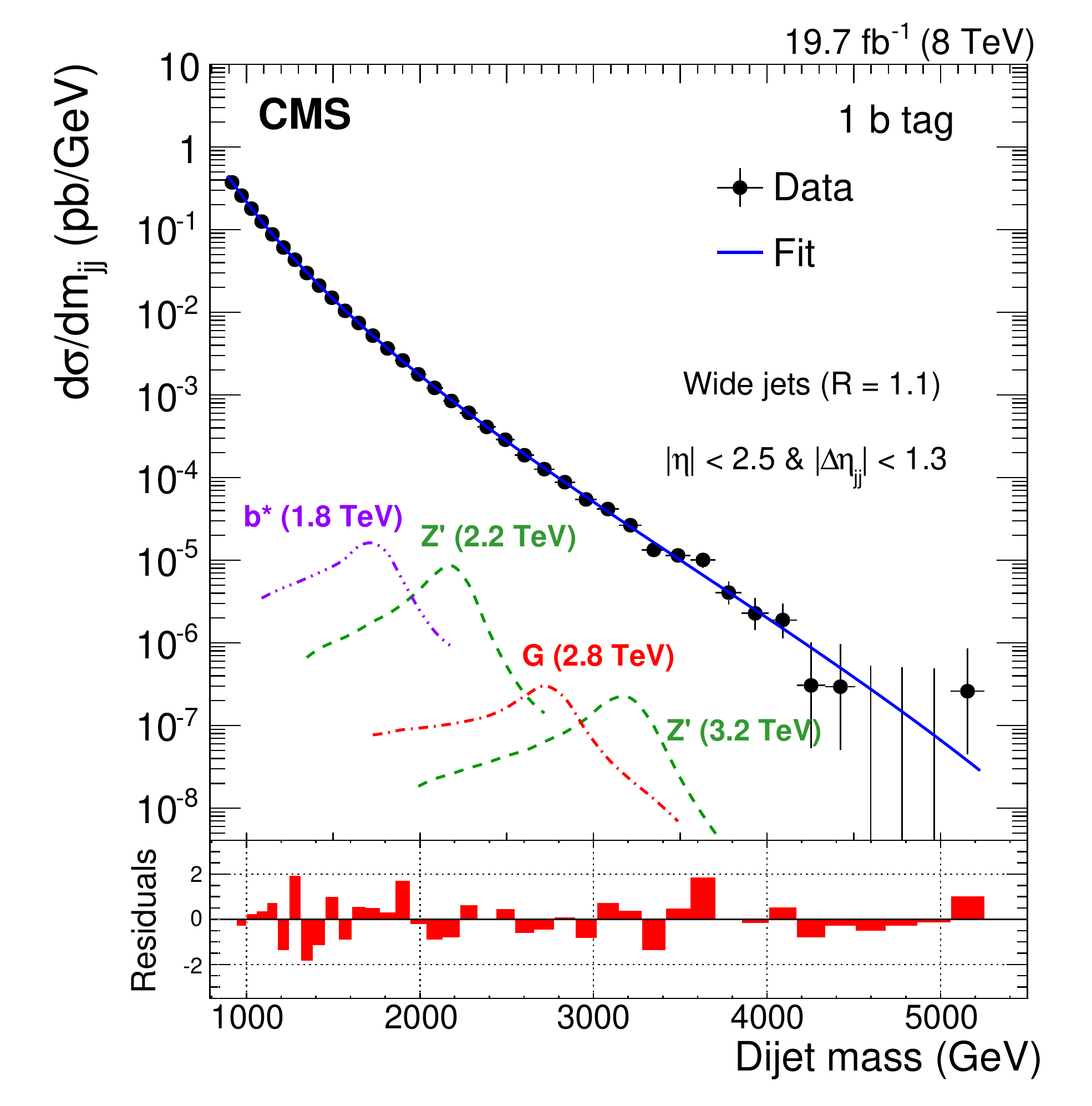}
    \includegraphics[width=0.48\textwidth]{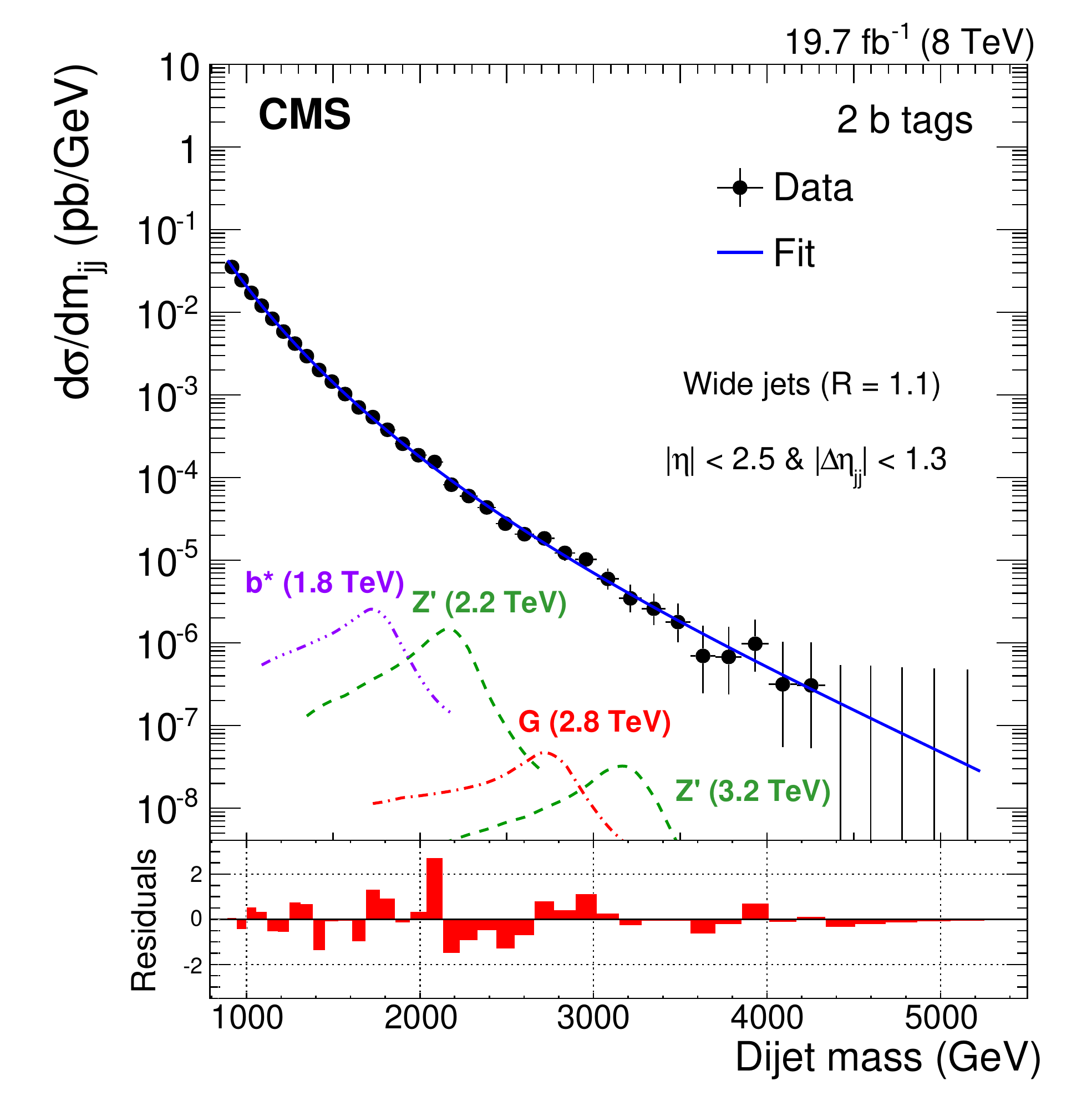}
    \caption{Dijet mass spectra (points) in different \PQb-tag
      multiplicity bins compared to a fit (solid curve).
      The vertical error bars are
      statistical only
      and the horizontal error bars are the bin widths.
      For comparison, signal distributions are shown for an excited \PQb quark of mass 1800\GeV, a \PZpr of mass 2200\GeV, an RS graviton of
      mass 2800\GeV, and a \PZpr of mass 3200\GeV.
      The bin-by-bin fit residuals scaled to the statistical uncertainty of the data,
      $(\text{data}-\text{fit})/\sigma_{\text{data}}$, are shown at the
      bottom of each plot.}
      \label{figDataAndMCbtag}
  \end{figure*}

A method based on data is used to estimate the background from multijet
production. We fit the following parameterization to the data:
\begin{equation}
\frac{\rd\sigma}{\rd\mjj}=
\frac{P_{0} (1 - x)^{P_{1}}}{x^{P_{2} + P_{3} \ln{(x)}}}
\label{eq:bkgfunction}
\end{equation}
with the variable $x=\mjj/\sqrt{s}$ and four free parameters
$P_0$, $P_1$, $P_2$, and $P_3$.  This functional form was used in
previous searches~\cite{Khachatryan:2010jd,PhysRevLett.106.029902,Chatrchyan:2011ns,CMS:2012yf,Chatrchyan:2013qhXX,ATLAS2010,Aad:2011aj,Aad:2011fq,ATLAS:2012pu,Aad:2014aqa,refCDFrun2}
to describe the distribution of both data and multijet background from
simulation. A Fisher $F$-test~\cite{Ftest} is used to confirm that no
additional parameters are needed to model these distributions for a
data set as large as the available one.  The fit of the data to the
function given in Eq.~(\ref{eq:bkgfunction}) returns a chi-squared
value of \chiSquare for \NDF degrees of freedom. The difference
between the data and the fit value is also shown at the bottom of
Fig.~\ref{figDataAndMC}, normalized to the statistical uncertainty of
the data. The 0\PQb, 1\PQb, and 2\PQb $\mjj$ dijet mass spectra are
shown in Fig.~\ref{figDataAndMCbtag}.  The function of
Eq.~(\ref{eq:bkgfunction}) is also fit to these data distributions.
The data are well described by this function and no significant
deviations from the background hypothesis are observed.

For comparison, signal distributions for
various narrow resonance models
are shown in both
Figs.~\ref{figDataAndMC} and \ref{figDataAndMCbtag}. These distributions
are obtained using \PYTHIA~{8.153}~\cite{Sjostrand:2007gs}, tune
4C~\cite{Corke:2010yf}, and the CMS detector simulation.

The $\PQq\PQq$ and $\Pg\Pg$ signal shapes are obtained from simulated samples of RS
graviton production, respectively $\PQq\PQq \to \cPG
\to \PQq\PQq$ and $\Pg\Pg \to \cPG \to
\Pg\Pg$. Graviton decays to all quark flavors other than top are
included; the top quark is excluded as its decays do not give rise to
the simple dijet experimental signature. The \PQq\Pg{} signal shapes are
obtained from simulations of excited quark production, $\PQq\Pg
\to \Qstar \to \PQq\Pg$. The simulated samples for
the inclusive analysis contain both \Ustar and \Dstar processes, while
for the \PQb-enriched analysis only \Bstar production is considered.
The predicted mass distributions have a Gaussian peak coming from the
jet energy resolution (JER) and a tail towards lower mass values induced by
the radiation of quarks and gluons at large angles. The contribution
of this low mass tail to the line shape depends on the parton content
of the resonance ($\Pq\Pq$, $\Pq\cPg$, or $\cPg\cPg$).  Resonances
containing gluons, which are more susceptible to
radiation than quarks, have a more pronounced tail. For high-mass
resonances, there is also another significant contribution depending
on both parton distributions and the natural width of the Breit--Wigner
resonance shape: when the resonance is produced by interaction of
non-valence partons in the proton, the low mass component of the
Breit--Wigner resonance shape is amplified by a larger parton
probability at low fractional momentum, producing a large tail at
low-mass values.

\section{Interpretation of the results}

Upper limits are set on the production cross section for different resonance
final states ($\PQq\PQq$, $\PQq\Pg$, $\Pg\Pg$, $\PQq\PQq$/$\PQb\PQb$, $\Pg\Pg$/$\PQb\PQb$, and $\PQb\Pg$) as a function of the resonance mass.
The limits are computed using a binned likelihood $L$ written as
a product of Poisson probability density functions
\begin{equation}
L = \prod_{i} \frac{\lambda_{i}^{n_i}\re^{-\lambda_i}}{n_{i}!},
\label{likelihood}
\end{equation}
where the product runs over the $\mjj$ bins.  For the $i${th}
$\mjj$ bin, $n_i$ is the observed number of events and
${\lambda_{i}} = {\mu}{N_{i}(S)} + {N_{i}(B)}$ denotes the expected
number of events. Here, $N_i(B)$ is the expected number of events from
multijet background, $N_i(S)$ is the expected number of signal events
for the benchmark models considered, and $\mu$ the ratio between the
signal production cross section and its corresponding benchmark value.
The background term $N_i(B)$ is estimated using the parameterization
of Eq.~(\ref{eq:bkgfunction}).

The dominant sources of systematic uncertainty are:
\begin{itemize}
\item uncertainty in the jet energy scale
  (JES)~\cite{1748-0221-6-11-P11002}, which translates into a
  \jecUncert relative uncertainty in the dijet mass, roughly
  independent of the mass value.  It is propagated to the search by
  shifting the reconstructed dijet mass for signal events by
  $\pm$\jecUncert;
\item uncertainty in the JER~\cite{1748-0221-6-11-P11002}, which translates
  into an uncertainty of 10\% in the dijet
  mass resolution~\cite{1748-0221-6-11-P11002}. This uncertainty is propagated
  to the search by increasing and decreasing by 10\% the reconstructed
  width of the dijet mass shape for the signal;
\item the precision in the overall normalization for the signal is
  limited by an uncertainty of \lumiUncert in the integrated
  luminosity~\cite{LUMIPASSummer2013};
\item \PQb tagging scale factors (${\sim}5\%$ for heavy and ${\sim}10\%$
  for light-flavor jets)~\cite{Chatrchyan:2012jua}, applied only in
  the dedicated \PQb-jet search.
\item uncertainties due to the choice of the background fit function are taken into account by the marginalization procedure described below.
\end{itemize}

Using studies based on simulations, the dependence of the signal
mass shapes on the number of pileup interactions is found to be
negligible. Similarly, no appreciable difference in the signal
acceptance is observed when different PDF sets are used.

For setting upper limits on signal cross sections
a Bayesian formalism~\cite{BayesianCowan} is used, with a uniform
prior for the signal cross section in the range
[0,$+\infty$]. For a given value of the resonance mass the data are fit to the
background function plus a signal line shape, the signal cross section
being a free parameter. The resulting fit function with the signal
cross section set to zero is used as the initial background
hypothesis. The uncertainty in the background shape is incorporated by
marginalizing over the background-fit parameters using uniform
priors. The integration is performed in a sufficiently large range
around the best-fit values such that the results are found to be
stable. Uncertainties due to alternative background fit functions are not
explicitly included since these variations are already covered by
the marginalization procedure with the default fit function (Eq.~\ref{eq:bkgfunction}).
Log-normal priors are used to model systematic uncertainties in the JES,  JER, integrated
luminosity, and \PQb-tagging efficiency, all treated as nuisance parameters.
The nuisance parameters are marginalized to derive a posterior
probability density function for the signal cross section.
The marginalization is performed using Markov chain Monte Carlo
integration implemented in the Bayesian Analysis Toolkit~\cite{Caldwell:2008fw}.

In the case of the search for $X \to \PQb\PQb$ resonances the
limit is obtained by combining the three event categories (0\PQb, 1\PQb, and 2\PQb).
The background distributions in the three samples are independently
varied in the fit.
The relative normalization of the signal distributions in the three
samples is determined by the ratio of the branching fractions of the $X$ resonance:
\begin{equation}
f_{\PQb\PQb} = \frac{\mathcal{B} (X\to \PQb\PQb)}{\mathcal{B} (X\to
   \mathrm{jj})}.
 \label{eq:bfrac}
 \end{equation}
As the fraction increases, events from a resonance in the 0\PQb
category shift into the 1\PQb and 2\PQb categories.
The distribution between the three categories also depends on the
tagging efficiencies shown in Fig.~\ref{fig:rate}. Mistags
of light-flavor jets are accounted for, according
to the quoted tagging probabilities.

\begin{figure*}[hbtp]
  \centering
    \includegraphics[width=0.48\textwidth]{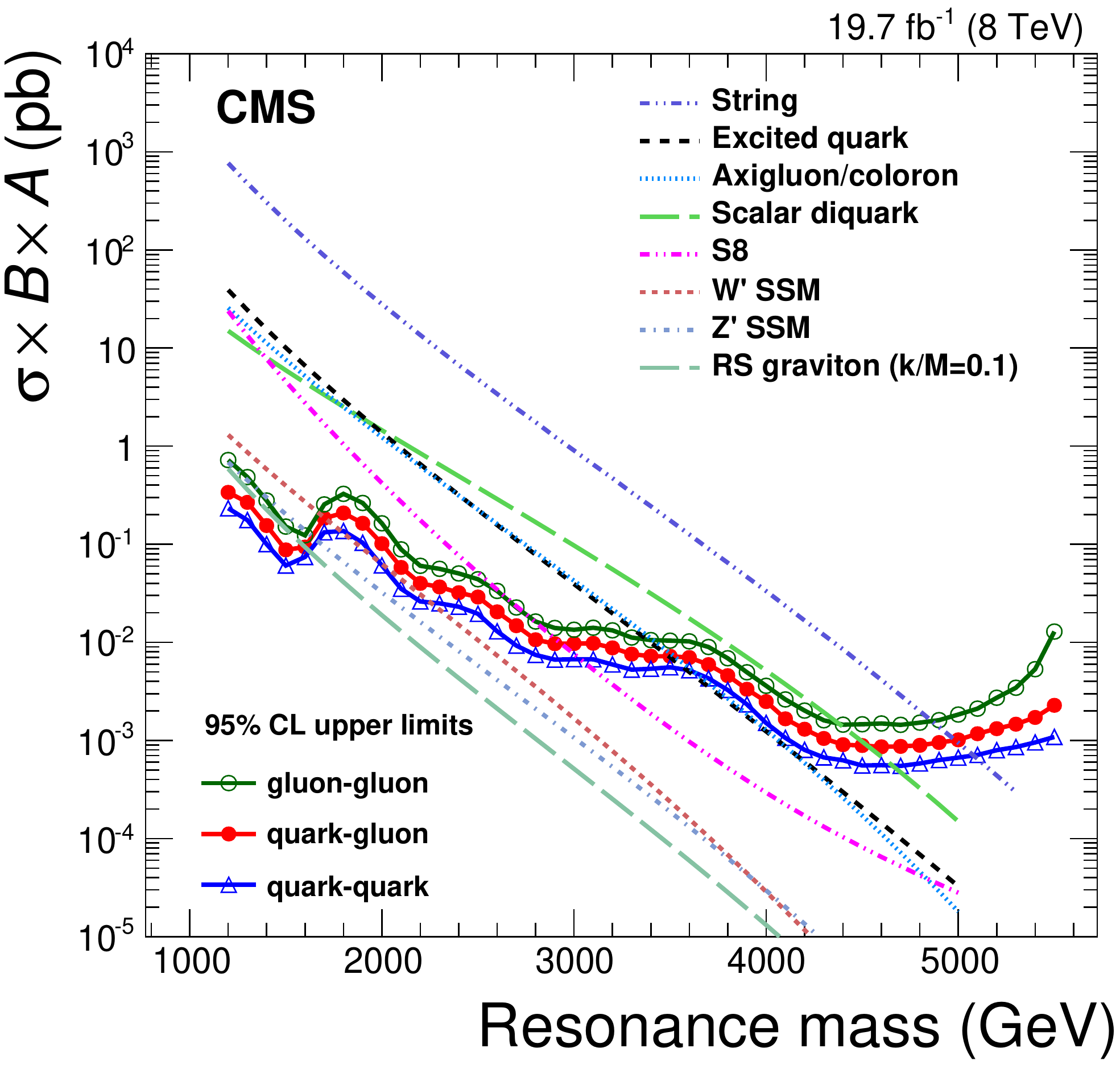}\\
    \includegraphics[width=0.48\textwidth]{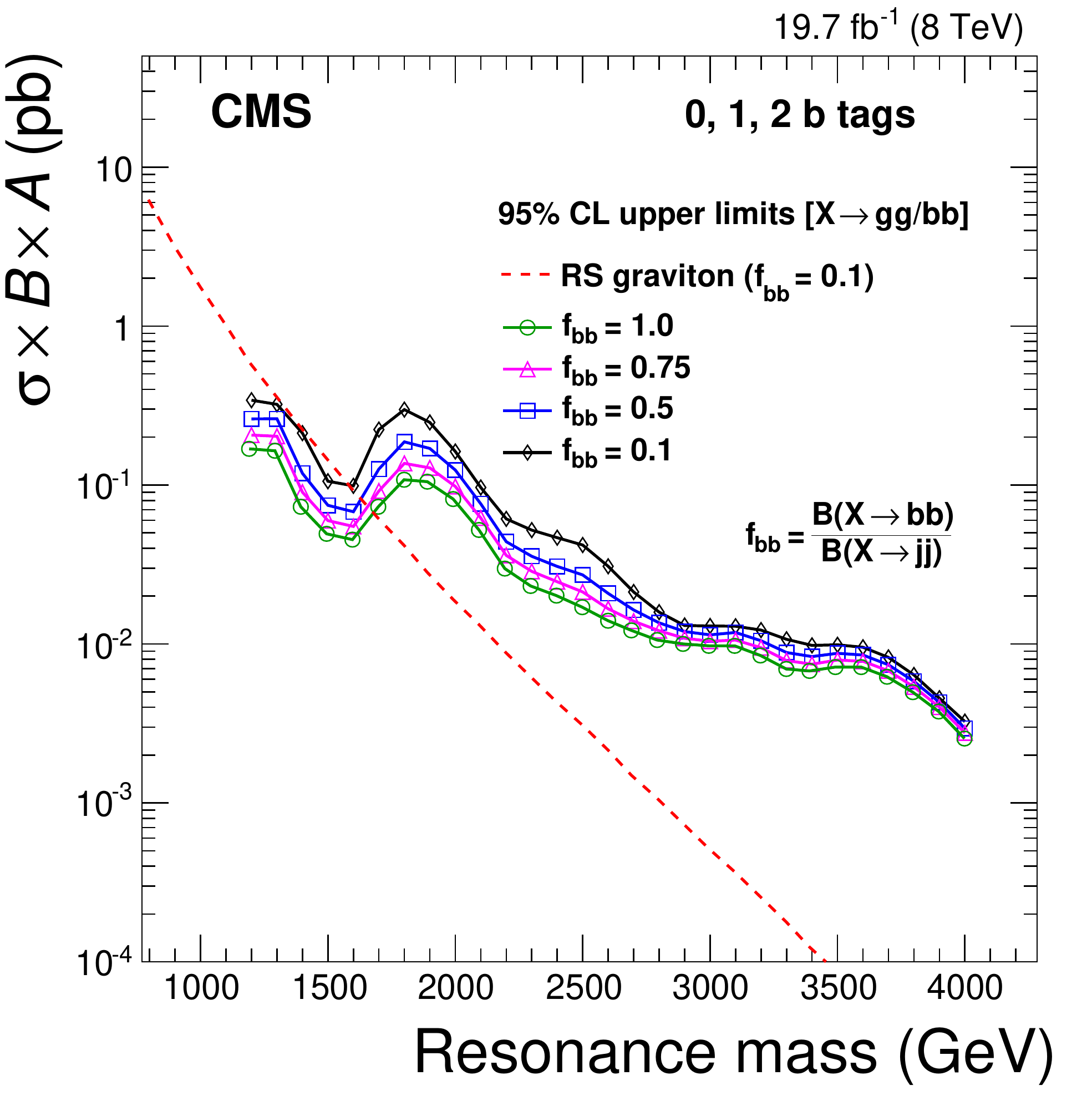}
    \includegraphics[width=0.48\textwidth]{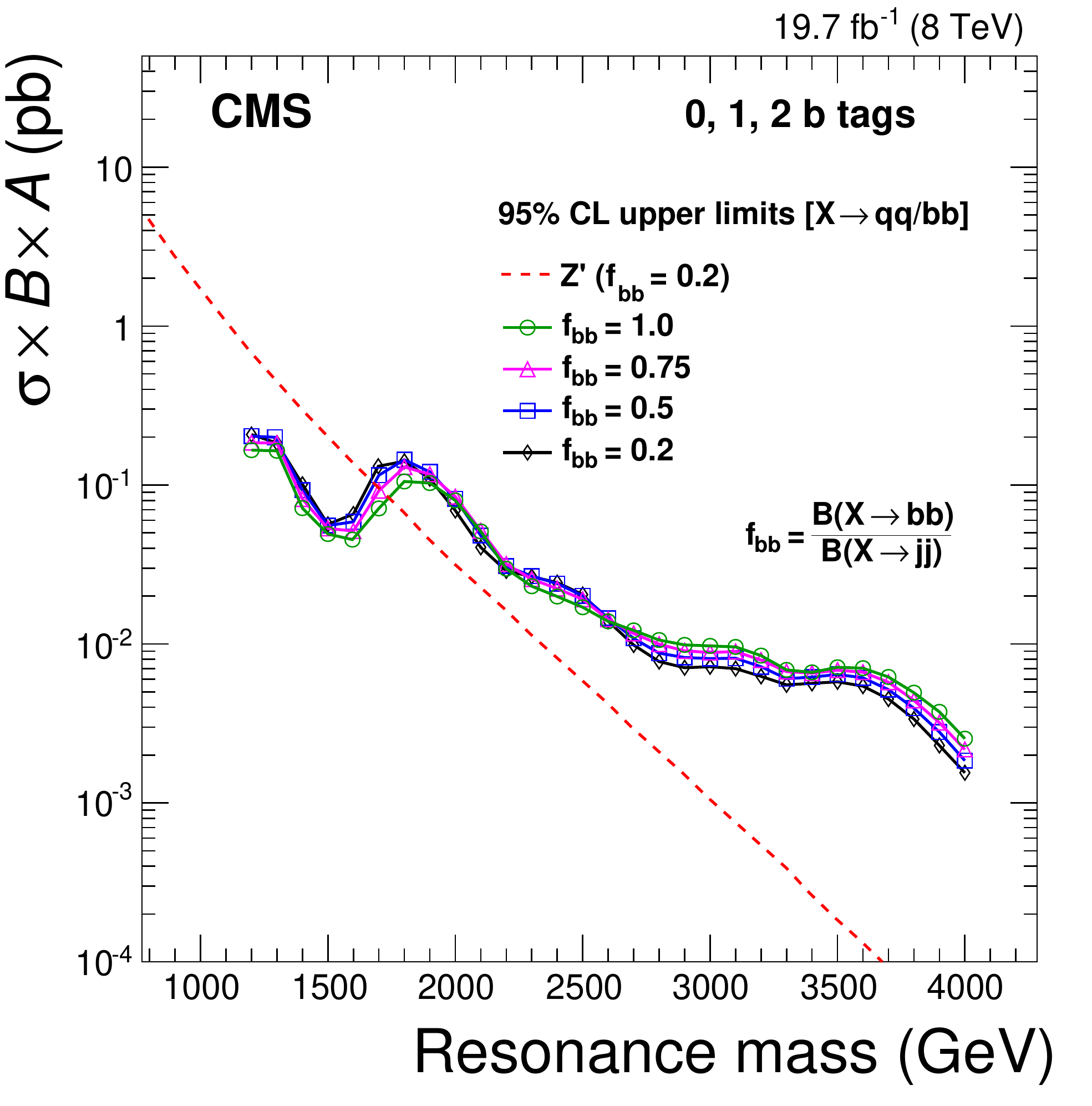}
    \caption{The observed 95\% \CL upper limits on \sba
      for narrow dijet resonances.  Top: limit on gluon-gluon,
      quark-gluon, and quark-quark narrow resonances from the inclusive
      analysis, compared to LO theoretical predictions for string
      resonances~\cite{Anchordoqui:2008di,Cullen:2000ef},
      excited quarks~\cite{ref_qstar,Baur:1989kv},
      axigluons~\cite{ref_axi,Bagger:1987fz,Chivukula:2011ng},
      colorons~\cite{ref_coloron},
      scalar diquarks~\cite{ref_diquark},
      S8 resonances~\cite{Han:2010rf},
      new SSM gauge bosons $\PWpr$ and
      $\cPZpr$~\cite{ref_gauge}, and RS
      gravitons~\cite{RS,ref_rsg,Bijnens:2001gh}.
Bottom left: combined limits on gg/bb
      resonances for different values of $f_{\PQb\PQb}$. The
      theoretical cross section for an RS graviton is
      shown for comparison.  Bottom right: combined limits on qq/bb resonances
      for different values of $f_{\PQb\PQb}$. The theoretical
      cross section for a $\cPZpr$ is
      shown for comparison.
      \label{figLimit}}
  \end{figure*}

Figure~\ref{figLimit} shows the observed model independent upper
limits at the 95\% confidence level (\CL) on
the product of the cross section
($\sigma$), the branching fraction into dijets ($\mathcal{B}$), and
the acceptance ($A$) for the kinematic requirements
described in Section~\ref{sec:EventSelection},
for narrow $\Pq\Pq$, $\cPg$, and $\cPg\cPg$ resonances. The acceptance
for isotropic decays is $A\approx 0.6$, independent of resonance
mass. The observed upper limits can be compared to LO predictions for
\sba at the parton level, without any detector
simulation, in order to determine mass limits on new particles.
The two partons in the LO process of the resonance decay should
both have pseudorapidity less than 2.5, their pseudorapidity
separation should be less than 1.3, and
their combined invariant mass should exceed 890\GeV.
The results shown are obtained in the narrow-width approximation using
CTEQ6L1 parton distributions~\cite{Pumplin:2002vw}.

\begin{figure*}[hbtp]
  \centering
    \includegraphics[width=\cmsFigWidth]{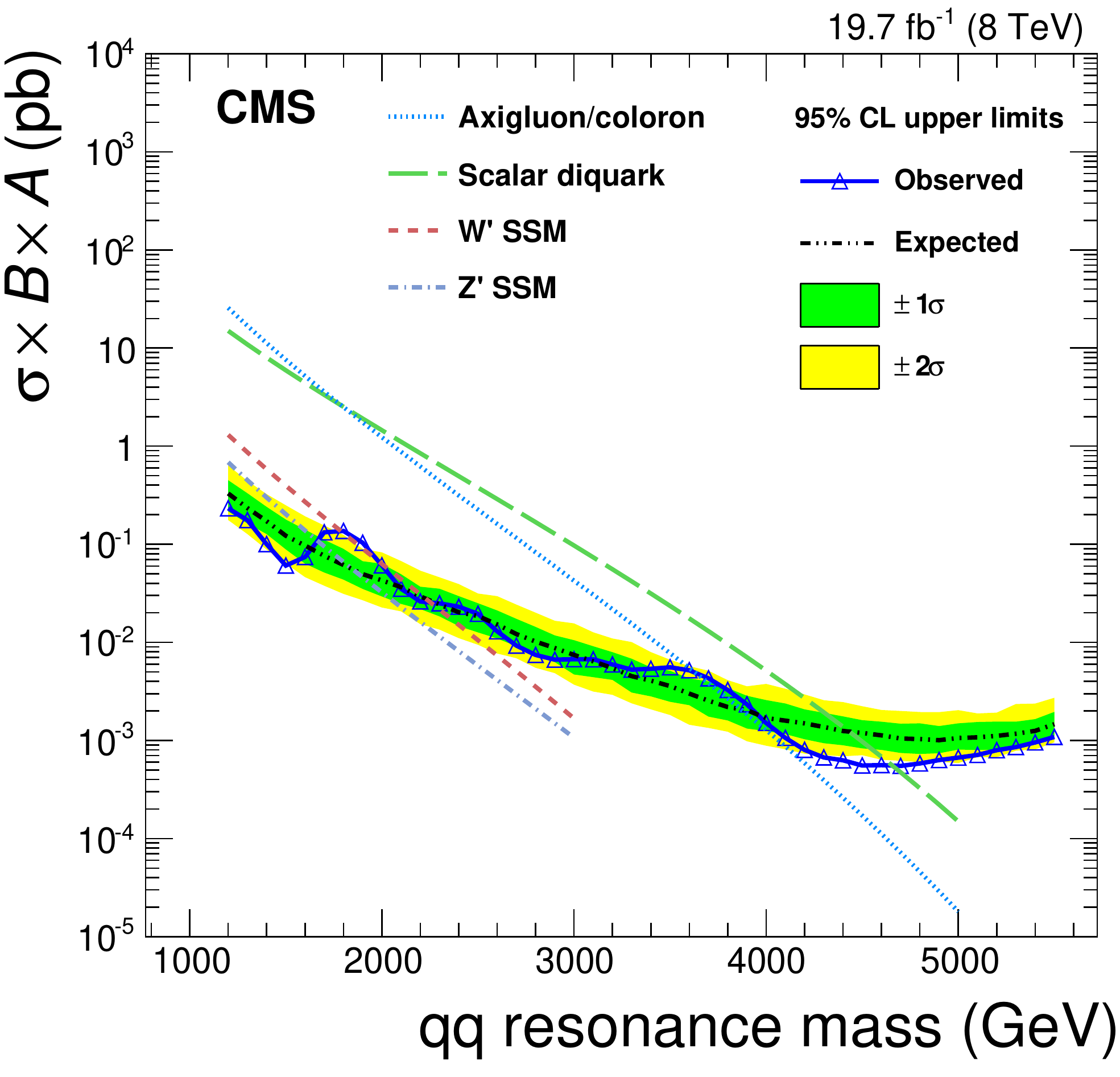}
    \includegraphics[width=\cmsFigWidth]{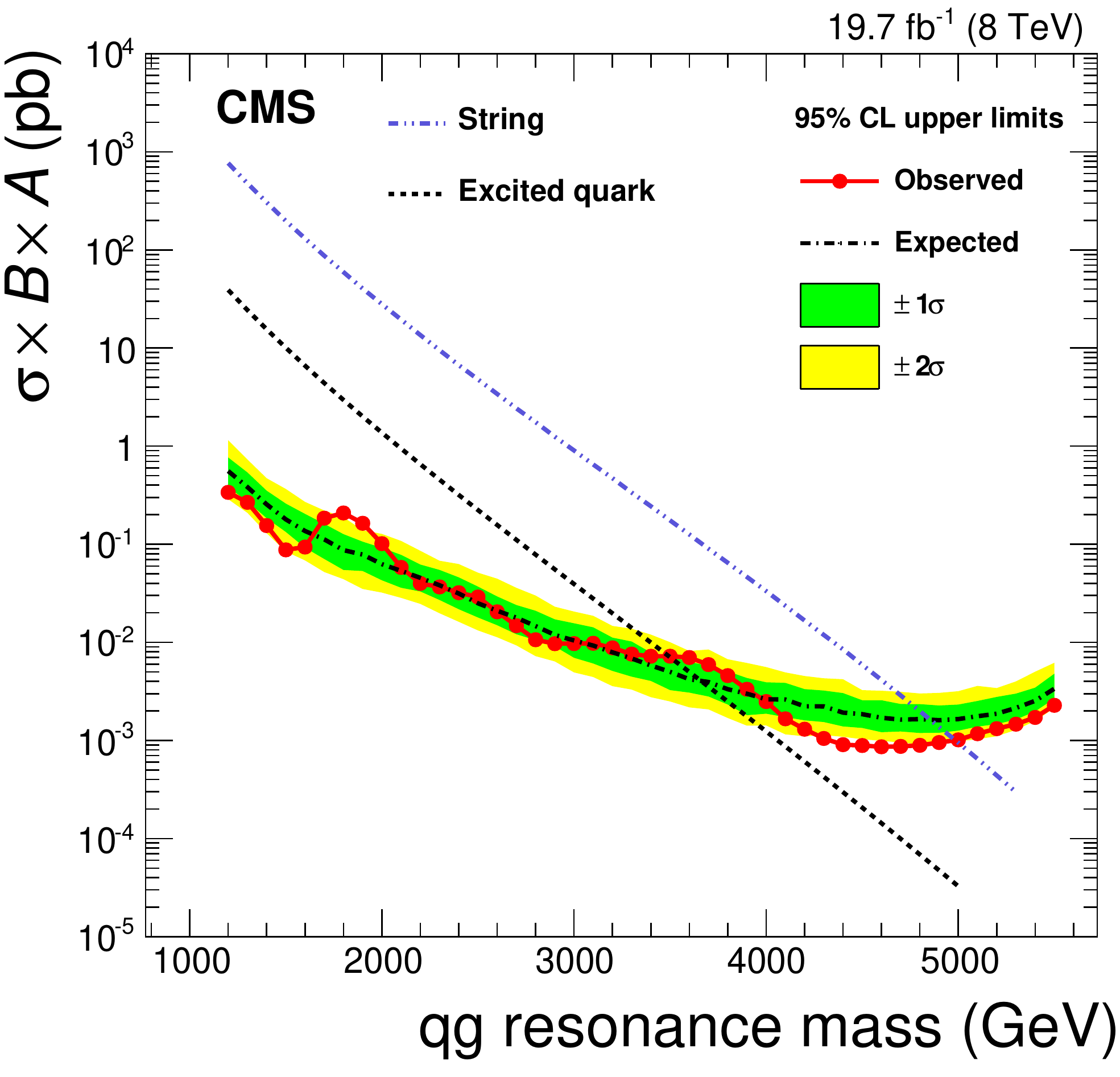}
    \includegraphics[width=\cmsFigWidth]{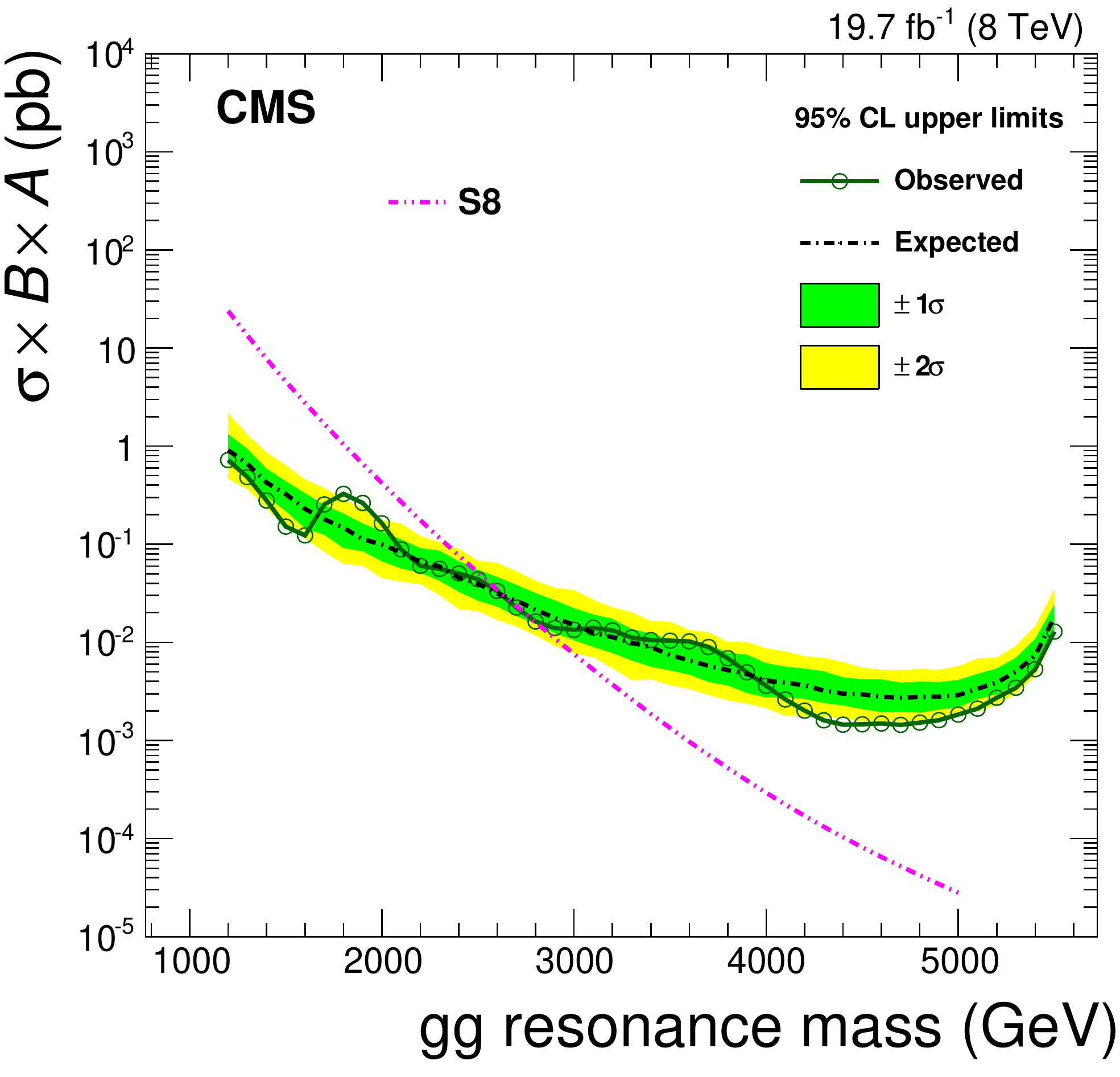}
    \includegraphics[width=\cmsFigWidth]{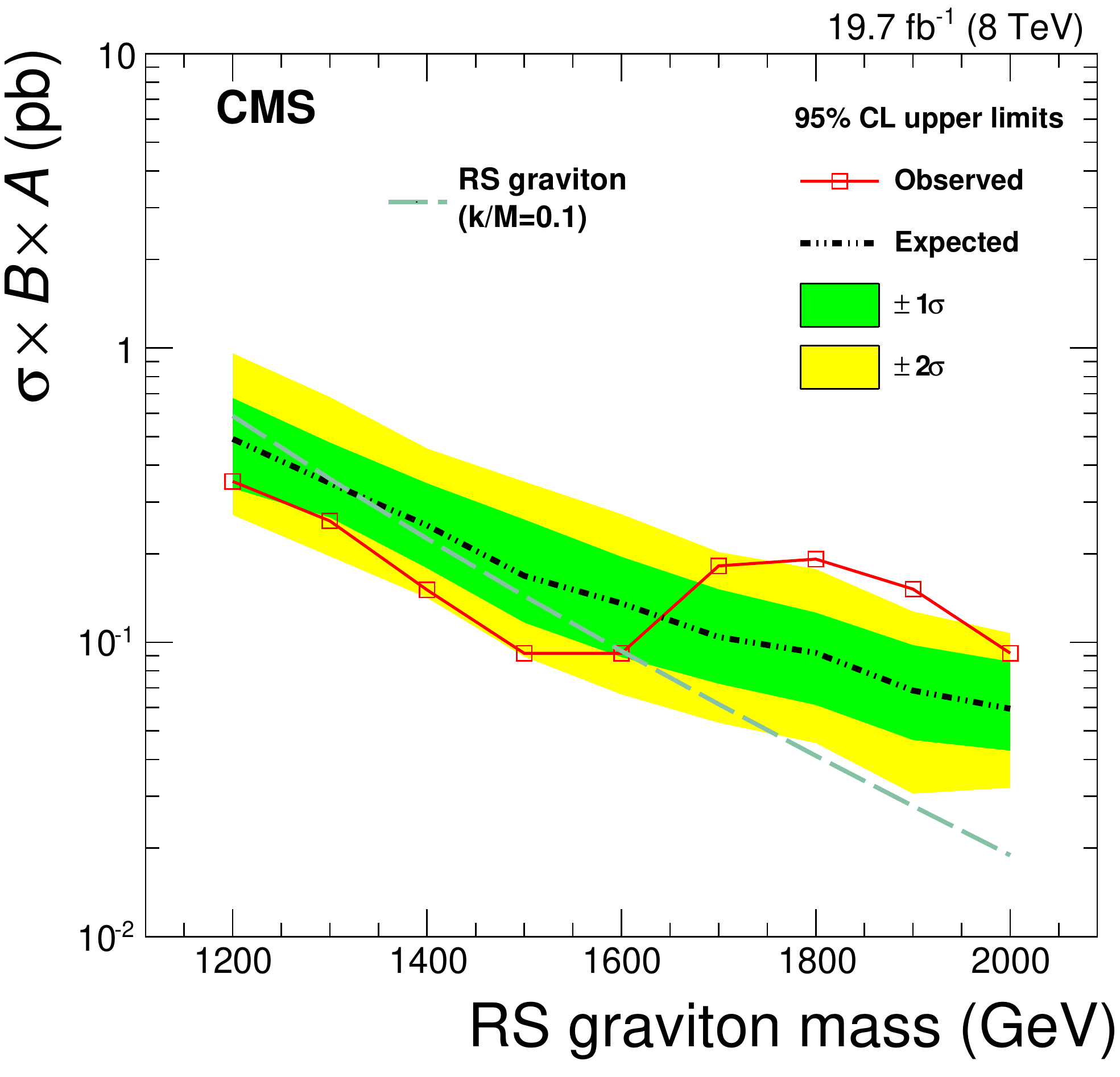}
    \caption{The observed 95\% \CL upper limits on \sba for narrow resonances decaying into $\Pq \Pq$ (top left), $\Pq\cPg$ (top right) and $\cPg\cPg$ (bottom left) final states,
      and for RS graviton resonances (bottom right). The limits are shown as
      points and solid lines. Also shown are the expected limits
      (dot-dashed dark lines) and their variation at the 1$\sigma$ and
      2$\sigma$ levels (shaded bands).  Predicted cross sections
      calculated at LO for various narrow resonances are also shown.
      \label{figExpectedLimit}}
  \end{figure*}

\begin{figure}[!htb]
  \centering
    \includegraphics[width=\cmsFigWidth]{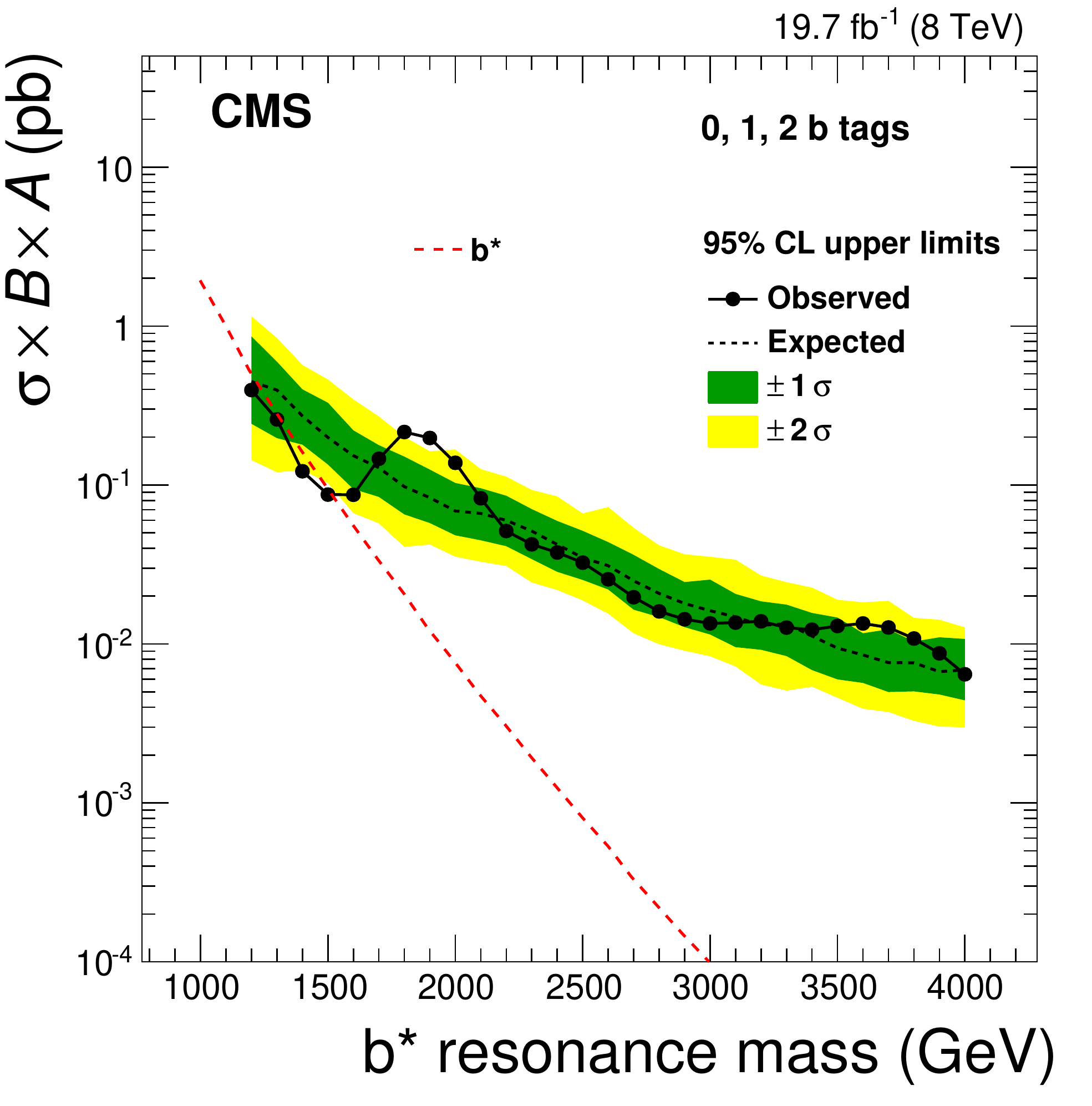}
    \caption{Observed and expected 95\% \CL upper limits on
      \sba with systematic uncertainties
      included, for
      $\Bstar \to\PQb\Pg $
      resonances, compared with the LO theoretical cross section for
      excited \PQb-quark production.
    }
    \label{fig:limits_obs_exp}
  \end{figure}

The expected limits on the cross section are estimated with
pseudo-experiments generated using background shapes, which are
obtained by signal-plus-background fits to the data.
Figure~\ref{figExpectedLimit} shows the expected limits and their
uncertainty bands for $\PQq\PQq$, $\PQq\Pg$, and $\Pg\Pg$ resonances compared to both observed limits and model predictions. For
the RS graviton, which couples either to a pair of gluons or to a
$\PQq\PQq$ pair, the model-dependent cross section limits are
obtained using a weighted average of the $\PQq\PQq$ (where $\PQq=\PQu$, \PQd, \PQc, \PQs, and \PQb,
excluding the top quark) and $\Pg\Pg$ dijet mass shapes. The weight factors
of about 0.5 correspond to the relative branching fractions for
these two final states derived from Ref.~\cite{Bijnens:2001gh}.
Figure~\ref{fig:limits_obs_exp} shows a similar plot for an excited \PQb quark.
The observed upper limits are reported in
Tables~\ref{tab:limits_obs_incl} and~\ref{tab:limits_obs_btag} for
the inclusive and \PQb-enriched analyses, respectively. The limits for
resonances with gluons in the final states are less restrictive than
those with quarks because the signal shapes are wider, as shown
for example in Figure~\ref{fig:shapes_broadResonance}.

\begin{table*}[!htbp]
 \topcaption{Observed 95\% \CL upper limits on \sba
   for narrow $\PQq\PQq$, $\PQq\Pg$, and $\Pg\Pg$ resonances, from the inclusive analysis
   for signal masses between 1.2 and 5.5\TeV.\label{tab:limits_obs_incl}}
 \centering
 \begin{scotch}{c*{3}{C}|c*{3}{C}}
   Mass   &  \multicolumn{3}{c|}{Upper limit on \sba\,(fb)} & Mass   &  \multicolumn{3}{c}{Upper limit on
     \sba\,(fb)}\\
\cline{2-4}\cline{6-8}
   (\TeVns)    & $\PQq\PQq$ & $\PQq\Pg$ & \multicolumn{1}{c|}{$\Pg\Pg$}
   & (\TeVns)  & $\PQq\PQq$ & $\PQq\Pg$ & $\Pg\Pg$ \\
   \hline
   1.2 & 230   & 340 & 720 & 3.4 & 5.4 & 7.3 & 11 \\
   1.3 & 180   & 270 & 480 & 3.5 & 5.6 & 7.2 & 10 \\
   1.4 & 100   & 160 & 280 & 3.6 & 5.2 & 7.0 & 10\\
   1.5 & 60 & 88 & 150 & 3.7 & 4.3 & 5.9 & 9.0 \\
   1.6 & 74 & 94 & 120 & 3.8 & 3.3 & 4.6 & 6.9 \\
   1.7 & 130   & 180 & 250 & 3.9 & 2.3 & 3.3 & 4.9 \\
   1.8 & 140   & 210 & 330 & 4.0 & 1.5 & 2.5 & 3.6 \\
   1.9 & 100   & 160  & 260 & 4.1 & 1.1 & 1.7& 2.6 \\
   2.0 & 60  & 100 & 160 & 4.2 & 0.80 & 1.3 & 2.0 \\
   2.1 & 35 & 58 & 89 & 4.3 & 0.67 & 1.0 & 1.6 \\
   2.2 & 26 & 40 & 60 & 4.4 & 0.63 & 0.91 & 1.4 \\
   2.3 & 25 & 37 & 56 & 4.5 & 0.56 & 0.89 & 1.5 \\
   2.4 & 23 & 32 & 50 & 4.6 & 0.56 & 0.86 & 1.5 \\
   2.5 & 20 & 29 & 44 & 4.7 & 0.55 & 0.87 & 1.4 \\
   2.6 & 13 & 20 & 33 & 4.8 & 0.58 & 0.89 & 1.5\\
   2.7 & 9.3 & 15 & 23 & 4.9 & 0.63 & 0.95 & 1.6\\
   2.8 & 7.4 & 11 & 16 & 5.0 & 0.67 & 1.0 & 1.8\\
   2.9 & 6.7 & 9.7 & 14 & 5.1 & 0.72 & 1.2 & 2.1 \\
   3.0 & 6.7 & 9.7 & 13 & 5.2 & 0.80 & 1.3 & 2.7 \\
   3.1 & 6.7 & 9.8 & 14 & 5.3 & 0.86 & 1.5 & 3.4\\
   3.2 & 5.9 & 8.8 & 13 & 5.4 & 0.95 & 1.7 & 5.3\\
   3.3 & 5.3 & 7.6 & 11 & 5.5 & 1.1 & 2.3 & 13\\
 \end{scotch}
\end{table*}

\begin{table*}[!htbp]
 \topcaption{Observed 95\% \CL upper limits on \sba
   for narrow $\Pg\Pg$/$\PQb\PQb$, $\PQq\PQq$/$\PQb\PQb$, and $\PQb\Pg$ resonances from the \PQb-enriched analysis,
   for signal masses between 1.2 and 4.0\TeV. The upper limits are given
 for different ratios $f_{\PQb\PQb}$ for $\Pg\Pg$/$\PQb\PQb$ and $\PQq\PQq$/$\PQb\PQb$ resonances, and for 100\% branching fraction into $\PQb\Pg$.\label{tab:limits_obs_btag}}
 \centering
 \begin{scotch}{c*{4}{C}c*{4}{C}}
   Mass   &  \multicolumn{9}{c}{Upper limit on \sba\,(fb)}\\
\cline{2-10}
   (\TeVns)  &  \multicolumn{4}{c|}{$\Pg\Pg$/$\PQb\PQb$} & \multicolumn{4}{c|}{$\PQq\PQq$/$\PQb\PQb$} &  $\PQb\Pg$\\
\cline{2-10}
   & $f_{\PQb\PQb}$=0.2 & $f_{\PQb\PQb}$=0.5  &  $f_{\PQb\PQb}$=0.75 &  $f_{\PQb\PQb}$=1.0 & $f_{\PQb\PQb}$=0.2 & $f_{\PQb\PQb}$=0.5  & $f_{\PQb\PQb}$=0.75 & $f_{\PQb\PQb}$=1.0 &  \\ \hline
1.2 &  340 &  260 &  210 &  170            &  210 &  200 &  180 &  170               &  400 \\
1.3 &  320 &  260 &  200 &  160    	       &  180 &  200 &  180 &  160    	    &  260 \\ 	   	
1.4 &  210 &  120 &  90 &  72           &  100 &  93 &  82 &  72            &  120 \\ 	   	
1.5 &  110 &  75 &  60 &  49          &  57 &  56 &  53 &  49          &  87 \\    	
1.6 &  99 &  68 &  55 &  45        &  66 &  59 &  51 &  45          &  87 \\    	
1.7 &  220 &  130 &  92 &  71          &  130 &  120 &  92 &  71            &  150 \\ 	   	
1.8 &  300 &  190 &  140 &  110    	       &  140 &  140 &  130 &  110    	    &  220 \\ 	   	
1.9 &  250 &  170 &  130 &  100    	       &  110 &  120 &  120 &  100    	    &  200 \\ 	   	
2.0 &  160 &  120 &  98 &  80           &  69 &  82 &  85 &  80         &  140 \\ 	   	
2.1 &  97 &  76 &  62 &  51        &  41 &  48 &  52 &  51          &  83 \\    	
2.2 &  61 &  44 &  36 &  30         &  29 &  31 &  32 &  30           &  51 \\    	
2.3 &  52 &  36 &  29 &  23        &  27 &  27 &  26 &  23          &  42 \\    	
2.4 &  47 &  31 &  25 &  20         &  24 &  24 &  22 &  20           &  38 \\    	
2.5 &  42 &  27 &  21 &  17        &  20 &  20 &  19 &  17            &  32 \\    	
2.6 &  31 &  21 &  17 &  14        &  14 &  15 &  14 &  14          &  26 \\    	
2.7 &  21 &  16 &  14 &  12        &  9.9 &  11 &  12 &  12         &  20 \\ 	   	
2.8 &  16 &  14 &  12 &  11        &  7.8 &  8.8 &  10 &  11         &  16 \\    	
2.9 &  13 &  12 &  11 &  9.9       &  7.1 &  8.2 &  9.0 &  9.9       &  14 \\    	
3.0 &  13 &  11 &  10 &  9.7        &  7.2 &  8.1 &  8.8 &  9.7      &  13 \\    	
3.1 &  13 &  12 &  11 &  9.6       &  7.0 &  8.1 &  9.0 &  9.6        &  14 \\    	
3.2 &  12 &  10 &  9.5 &  8.4       &  6.3 &  7.2 &  7.9 &  8.4      &  14 \\    	
3.3 &  11 &  8.8 &  7.9 &  6.9     &  5.5 &  6.0 &  6.7 &  6.9       &  13 \\    	
3.4 &  9.8 &  8.3 &  7.5 &  6.6    &  5.7 &  6.2 &  6.6 &  6.6      &  12 \\    	
3.5 &  9.9 &  8.7 &  7.9 &  7.1    &  5.8 &  6.4 &  6.8 &  7.1      &  13 \\    	
3.6 &  9.5 &  8.5 &  7.8 &  7.0     &  5.4 &  6.1 &  6.6 &  7.0       &  13 \\    	
3.7 &  8.2 &  7.4 &  6.8 &  6.2    &  4.5 &  5.2 &  5.7 &  6.2      &  13 \\    	
3.8 &  6.4 &  5.8 &  5.4 &  4.9    &  3.4 &  3.9 &  4.4 &  4.9      &  11 \\    	
3.9 &  4.6 &  4.3 &  4.0 &  3.7     &  2.3 &  2.8 &  3.2 &  3.7      &  8.7 \\   	
4.0 &  3.2 &  2.9 &  2.8 &  2.5    &  1.5 &  1.8 &  2.2 &  2.5      &  6.4 \\   	
 \end{scotch}
\end{table*}

New particles are excluded at 95\% \CL in mass regions for which the
theoretical curve lies above the observed upper limit for the
appropriate final state.  The observed and expected mass exclusions
for various models are reported in Table~\ref{exclusionLimits_fat}.
Table~\ref{exclusionLimits_fat} also shows limits on
axigluons/colorons and S8 resonances, interpreted as wide resonances,
as discussed in the next section.  For comparison with previous
searches, we quote here the new limits at a 95\% \CL on these two models
interpreted as narrow resonances, as shown in Fig.~\ref{figExpectedLimit}. These
limits provide reference values to quantify the impact of a
non-negligible resonance width. For narrow axigluons/colorons the
observed and expected mass limits are 3.7 and 3.9\TeV,
respectively.  The corresponding exclusion limits for narrow S8
resonances are 2.7 and 2.6\TeV, respectively.

\begin{table*}[htb]
  \topcaption{Observed and expected 95\% \CL exclusions on the mass of
    various resonances. Systematic uncertainties are taken into
    account. For excited \PQb quark the expected mass limit is below the
    range of this analysis. For the Axigluon/coloron and
    color-octet scalar only observed mass limits are computed.
    \label{exclusionLimits_fat}}
  \centering
  \begin{scotch}{cccc}
    \multicolumn{4}{c}{Inclusive search} \\
    \hline
    Model & Final state & Observed mass & Expected mass \\
               & & exclusion (\TeVns)& exclusion (\TeVns)\\
    \hline
    String resonance (S) & $\PQq\Pg$  & [1.2,5.0]  & [1.2,4.9] \\
    Excited quark (\Qstar) & $\PQq\Pg$  & [1.2,3.5]  & [1.2,3.7] \\
    $E_{6}$ diquark (D) & $\PQq\PQq$  & [1.2,4.7]  & [1.2,4.4] \\
    \PWpr boson (\PWpr) & $\PQq\PAQq$  & [1.2,1.9] $+$ [2.0,2.2]  & [1.2,2.2] \\
    \PZpr boson (\PZpr) & $\PQq\PAQq$  & [1.2,1.7] & [1.2,1.8] \\
    RS graviton ($\cPG$), $k/\overline{M}_\text{Pl}=0.1$ & $\PQq\PAQq + \Pg\Pg$ & [1.2,1.6]  & [1.2,1.3] \\
    \hline
    \multicolumn{4}{c}{\PQb-enriched search} \\
    \hline
    Excited \PQb quark (\Bstar) & $\PQb\Pg$  & [1.2,1.6]  &  \\
    \hline
    \multicolumn{4}{c}{Wide resonance search} \\
    \hline
    Axigluon (A)/coloron (C) & $\PQq\PAQq$  & [1.3,3.6] &  \\
    Color-octet scalar (S8) & $\Pg\Pg$ & [1.3,2.5] &  \\
  \end{scotch}
\end{table*}

\section{Implications for wide resonances} \label{sec:WideRes}

In the previous sections we have described a search for narrow dijet
resonances, where the intrinsic resonance width is negligible
compared to the experimental dijet mass resolution.
In order to quantify the impact of this search on models with wide
resonances, we consider the case of an RS graviton produced via
$\Pq\Pq$ and $\cPg \cPg$ annihilation and decaying, respectively, to $\Pq
\Pq$ and $\cPg \cPg$ final states.
Samples are generated with \PYTHIA scanning the plane defined
by the graviton mass $M$ and the coupling parameter $k/\overline{M}_\text{Pl}$.
For resonances with mass at the TeV scale, the width-to-mass ratio of
the resonance is $\Gamma/M \approx 1.4
(k/\overline{M}_\text{Pl})^2$~\cite{Bijnens:2001gh}.
The excluded signal cross section is presented as a function of the resonance mass and width,
separately for the $\Pq\Pq$ and $\cPg\cPg$ final states,
in order to allow the interpretation of the results in a generic
model.

\begin{figure}[!htb]
 \centering
 \includegraphics[width=\cmsFigWidth]{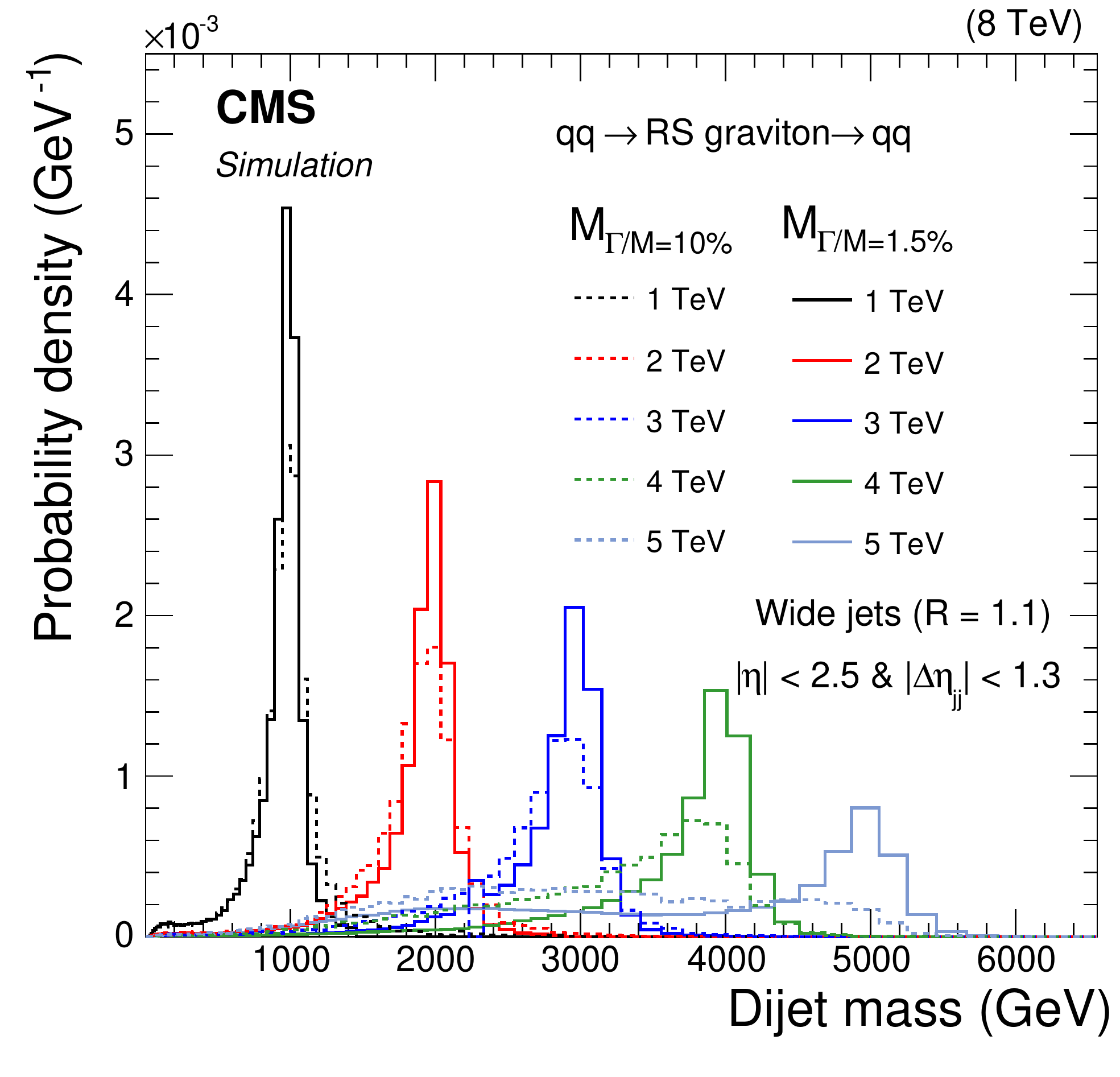}
 \includegraphics[width=\cmsFigWidth]{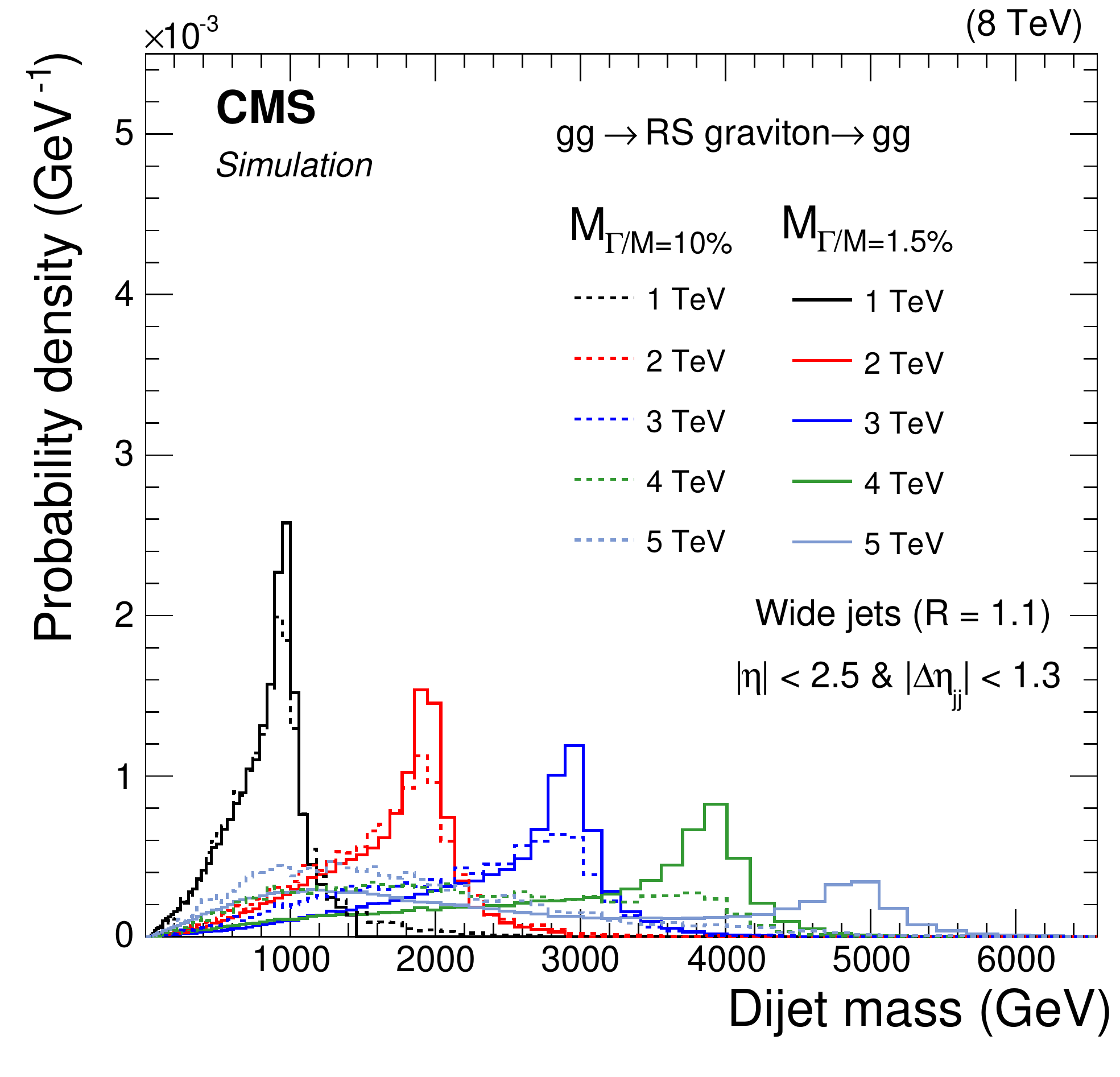}
 \caption{Dijet mass distributions for $\PQq\PQq$ (\cmsLeft) and $\Pg\Pg$ (\cmsRight)
   resonances with masses of 1, 2, 3, 4, and 5\TeV and two different
   values of $\Gamma/M$ (10\% and 1.5\%).
The corrections for the difference in the JES between a parametric simulation
and the \GEANTfour-based CMS simulation have been applied.
\label{fig:shapes_broadResonance}}
\end{figure}

\begin{figure}[!htb]
 \centering
 \includegraphics[width=\cmsFigWidth]{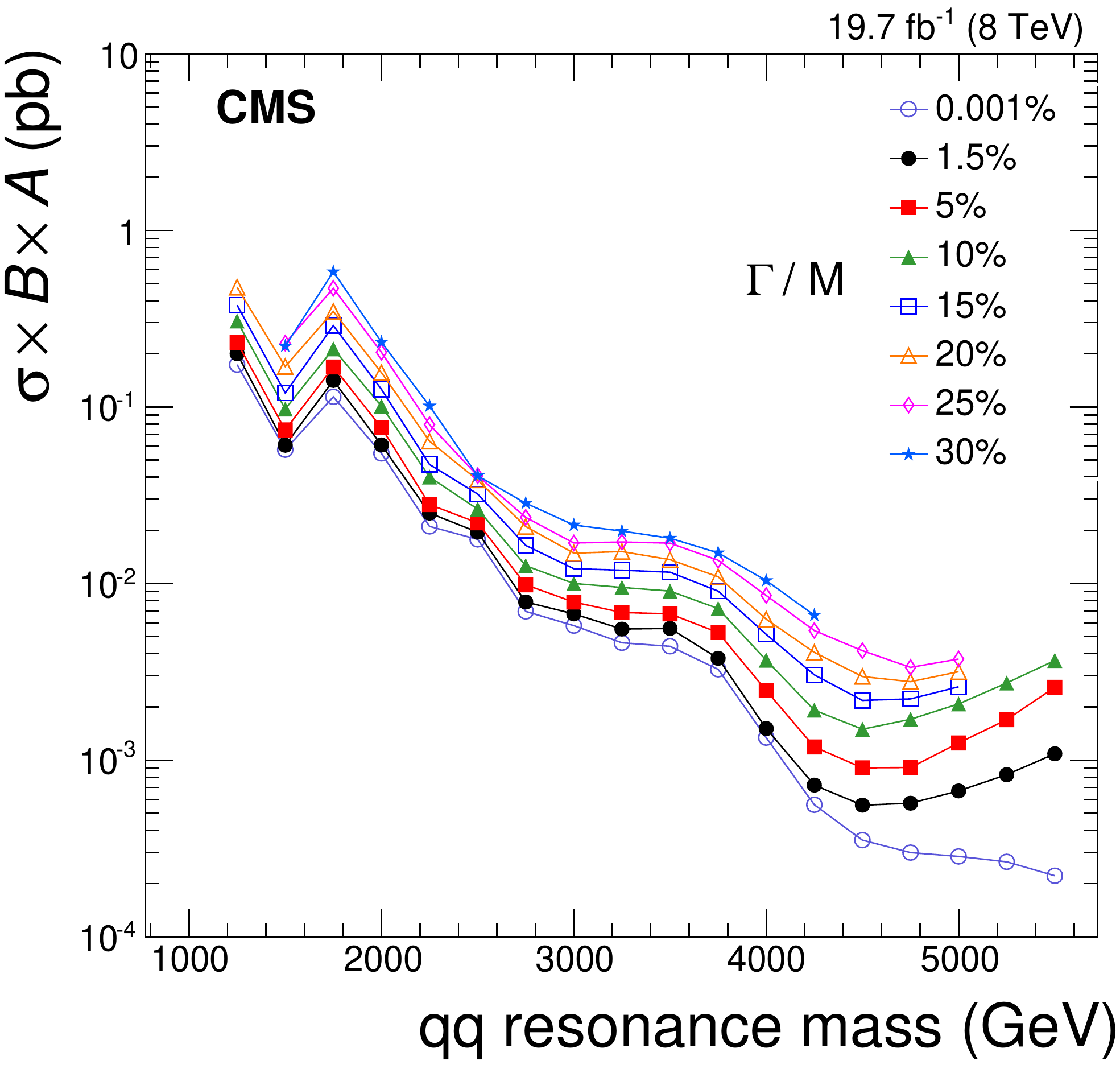}
 \includegraphics[width=\cmsFigWidth]{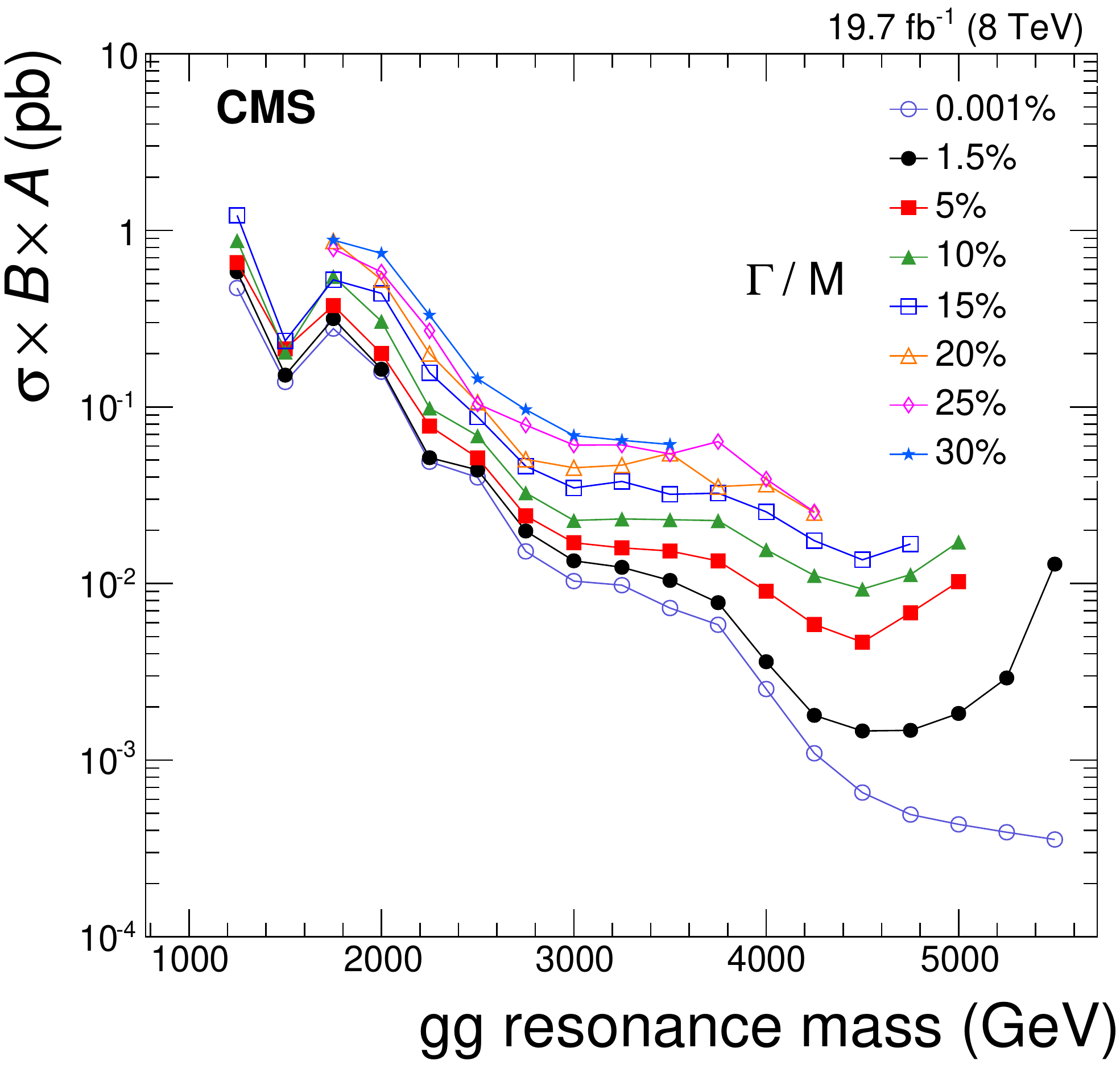}
 \caption{Observed 95\% \CL upper limits on \sba as a function of the resonance mass for
   different values of the width-to-mass ratio $\Gamma/M$, computed
   for $\Pq \Pq \to \cPG \to \Pq \Pq$ (\cmsLeft) and $\cPg \cPg \to \cPG \to \cPg \cPg$
   (\cmsRight).\label{fig:limit_broadResonance}}
\end{figure}

The ($M$, $k/\overline{M}_\text{Pl}$) scan is performed using events
generated with \PYTHIA~{8.153} and a parametric, fast simulation of the
CMS detector~\cite{FastSim}. The predicted signal distribution is
corrected for the difference in the JES between the fast simulation
and the \GEANTfour-based CMS
simulation. Figure~\ref{fig:shapes_broadResonance} shows the corrected
dijet mass distributions for several different values of resonance mass
$M$ and width-to-mass ratio $\Gamma/M$.  The excluded cross section
at 95\% \CL as a function of the resonance mass is shown in
Fig.~\ref{fig:limit_broadResonance} for different values of
$\Gamma/M$.
At resonance masses around 1--2\TeV, the value
of the excluded cross section slightly increases with
the resonance width, as expected from the gradual widening of
the core of the resonance approximately independent of the tail.
For large resonance masses, the exclusion limit for wide resonances is
worse than the narrow resonance limits by at least one order of
magnitude.  This different behavior is caused by the enhancement in the
low-mass tail of the dijet mass signal shape from partons with low
fractional momentum, which is more important for high-mass resonances.
Nevertheless, the analysis remains sensitive to new resonances up to
$\Gamma/M \approx 30\%$.  The cross section limits are reported in
Table~\ref{tabggfastLimit_newrange} for $\PQq\PQq$ and $\Pg\Pg$ final states.  The
limits are quoted for a range of masses and widths that satisfies two
conditions: (i) at low resonance mass, the core of the signal shape is preserved
after the trigger selection $\mjj>\minMjjCut\GeV$, (ii) at
high resonance mass, the presence of the low-mass tails in the signal shape does
not significantly affect the limit value.
This latter condition is enforced by
requiring that the expected limit derived for a truncated signal shape is
close to that derived for the full shape, within the typical uncertainty of
30\% in the expected limits. The truncated shape is cut off at 85\% of the
nominal resonance mass, and the corresponding limit corrected for the
difference in acceptance because of the truncation.

\begin{figure}[!htb]
 \centering
 \includegraphics[width=\cmsFigWidth]{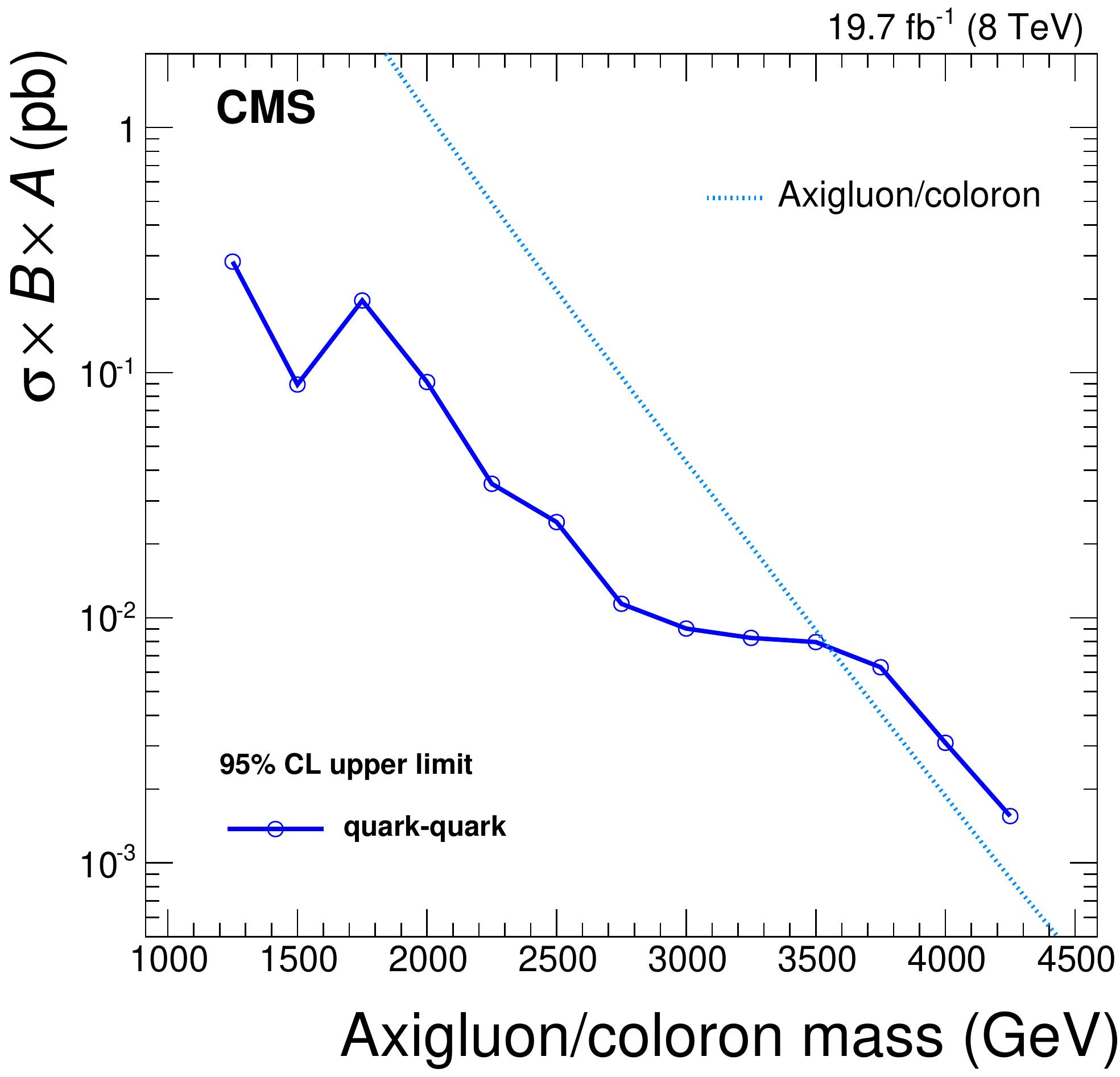}
 \includegraphics[width=\cmsFigWidth]{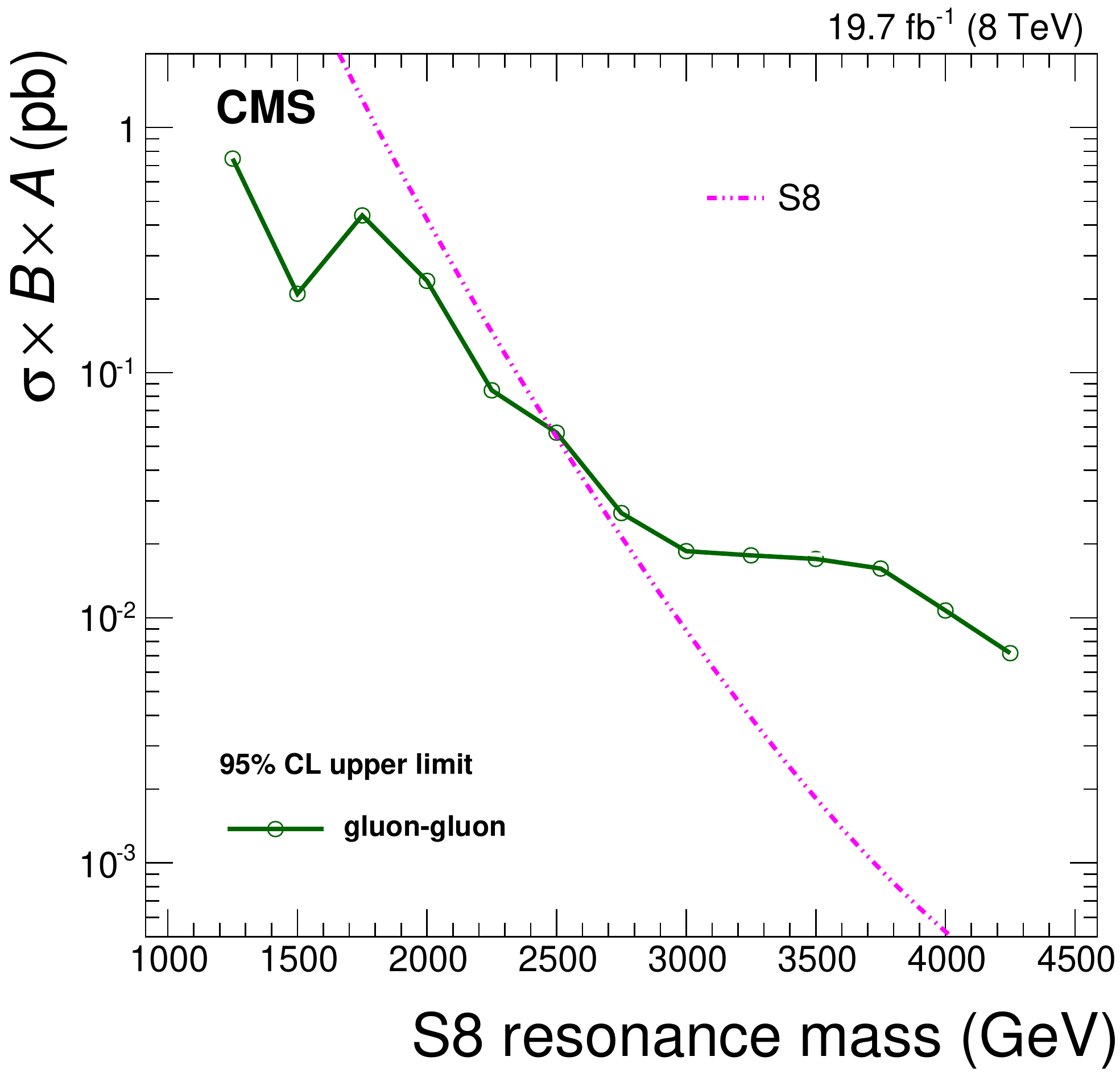}
 \caption{Observed 95\% \CL upper limits on
      \sba with systematic uncertainties
      included for axigluon/coloron (\cmsLeft) and S8 (\cmsRight) wide resonances,
compared to the corresponding theoretical predictions.
The axigluon/coloron and S8 resonances have a
relative width $\Gamma/M$ between 5\% and 10\%.
      More details on the cross section calculations for wide resonances are reported
      in the Appendix.
      \label{fig:limit_axigluon_S8}}
\end{figure}

We present below an example, illustrating how to use these generic
upper limits on the cross section to set lower mass limits for
specific models of wide resonances.  The axigluon/coloron and S8
resonances represent good benchmark models for this study, having
relative widths $\Gamma/M$ equal to $\alpha_S$ and $5/6
\alpha_S$, respectively (where $\alpha_S$ is the SM strong coupling
evaluated at an energy scale equal to the resonance mass). $\Gamma/M$
is therefore between 5\% and 10\%, slightly decreasing with the
increase in the resonance mass because of the running of the strong
coupling constant.  New cross section upper limits for
axigluon/coloron and S8 resonances are produced, which are,
respectively, a linear interpolation between the $\Gamma/M=5\%$ and
10\% $\PQq\PQq$ and $\Pg\Pg$ limits reported in
Table~\ref{tabggfastLimit_newrange}. The resulting cross section upper
limits are shown in Fig.~\ref{fig:limit_axigluon_S8} where they are
compared to theoretical predictions to extract the lower mass limits
on axigluon/coloron and S8 resonances reported in
Table~\ref{exclusionLimits_fat}. More details on the cross section
calculations for wide resonances are reported in the Appendix.

\begin{table}[htb]
  \centering
\topcaption{Observed 95\% \CL upper limits on \sba as a function of resonance mass for
   several values of the width-to-mass ratio $\Gamma/M$, computed
   for $\Pq \Pq \to \cPG \to \Pq \Pq$ and $\cPg \cPg \to \cPG \to \cPg
   \cPg$. The missing entries correspond to the region where the two conditions
for the validity of the wide resonance analysis are not satisfied (see text).}
\begin{scotch}{ccccccccc}
\multicolumn{9}{c}{Observed 95\% \CL $ \sba$ limit (fb)} \\
\hline
Mass   &  \multicolumn{8}{c}{$\Gamma/M$ (\%) for $\Pq \Pq \to \cPG \to \Pq \Pq$} \\
\cline{2-9}
 (\TeV)  &  0.001  &  1.5  &  5 & 10 & 15 & 20 & 25 & 30\\
\hline
1.25 & 170 & 200 & 230 & 310 & 380 & 470 &  & \\
1.50 & 57 & 61 & 74 & 97 & 120 & 170 & 230  & 220 \\
1.75 & 110 & 140 & 170 & 210 & 290 & 350 & 470 & 580 \\
2.00 & 55 & 61 & 76 & 100 & 130 & 160 & 200 & 230 \\
2.25 & 21 & 25 & 28 & 40 & 47 & 64 & 80 & 100 \\
2.50 & 18 & 20 & 22 & 26 & 32 & 39 & 41 & 41 \\
2.75 & 6.9 & 7.9 & 9.8 & 13 & 16 & 21 & 24 & 28 \\
3.00 & 5.8 & 6.7 & 7.8 & 10 & 12 & 15 & 17 & 21 \\
3.25 & 4.6 & 5.5 & 6.9 & 9.5 & 12 & 15 & 17 & 20 \\
3.50 & 4.4 & 5.6 & 6.7 & 9.1 & 12  & 14 & 17 & 18 \\
3.75 & 3.2 & 3.8 & 5.3 & 7.2 & 9.1 & 11 & 14 & 15 \\
4.00 & 1.3 & 1.5 & 2.5 & 3.7 & 5.2 & 6.3 & 8.6 & 10 \\
4.25 & 0.56 & 0.72 & 1.2 & 1.9  & 3.0 & 4.1 & 5.4 & 6.6 \\
4.50 & 0.35 & 0.56 & 0.90 & 1.5 & 2.2 & 3.0 & 4.2 &   \\
4.75 & 0.30 & 0.57 & 0.91 & 1.7 & 2.2  & 2.8 & 3.3 &   \\
5.00 & 0.28 & 0.67 & 1.2 & 2.1 & 2.6 & 3.2 & 3.7 &   \\
5.25 & 0.26 & 0.83 & 1.7 & 2.7 &   &  &  &  \\
5.50 & 0.22 & 1.1 & 2.6 & 3.7 &  &  &  &  \\
\hline
\multicolumn{9}{c}{Observed 95\% \CL $ \sba$ limit (fb)} \\
\hline
Mass   &  \multicolumn{8}{c}{$\Gamma/M$ (\%) for $\cPg \cPg \to \cPG \to \cPg \cPg$} \\
\cline{2-9}
(\TeV)  &  0.001  &  1.5  &  5 & 10 & 15 & 20 & 25 & 30\\
\hline
1.25 & 470 & 580 & 660 & 880 & 1200 &  &  &  \\
1.50 & 140 & 150 & 210 & 200 & 240 &  &  &  \\
1.75 & 280 & 320 & 370 & 550 & 520 & 870 & 780 & 880 \\
2.00 & 160 & 160 & 200 & 310 & 440 & 530 & 580 & 740 \\
2.25 & 49 & 52 & 78 & 99 & 160 & 200 & 270 & 330 \\
2.50 & 40 & 44 & 51 & 69 & 88 & 110 & 100 & 140 \\
2.75 & 15 & 20 & 24 & 33 & 46 & 50 & 79 & 97 \\
3.00 & 10 & 13 & 17 & 23 & 35 & 45 & 61 & 69 \\
3.25 & 9.8 & 12 & 16 & 23 & 38 & 47 & 61 & 65 \\
3.50 & 7.2 & 10 & 15 & 23 & 32  & 55 & 54 & 61 \\
3.75 & 5.8 & 7.8 & 13 & 23 & 32 & 35 & 64 &  \\
4.00 & 2.5 & 3.6 & 9.0 & 16 & 25 & 36 & 40 &  \\
4.25 & 1.1 & 1.8 & 5.9 & 11  & 17 & 25 & 25 &  \\
4.50 & 0.65 & 1.5 & 4.7 & 9.3 & 14 &  &  &  \\
4.75 & 0.49 & 1.5 & 6.8 & 11 & 17  &  &  & \\
5.00 & 0.43 & 1.8 & 10 & 17 &  &  &   &   \\
5.25 & 0.39 & 2.9 &  &  &  &  &  &  \\
5.50 & 0.36 & 1.3 &  &  &  &  &  &  \\
\end{scotch}
\label{tabggfastLimit_newrange}
\end{table}

\section{Implications for quantum black holes}

\begin{figure}[!htb]
 \centering
 \includegraphics[width=\cmsFigWidth]{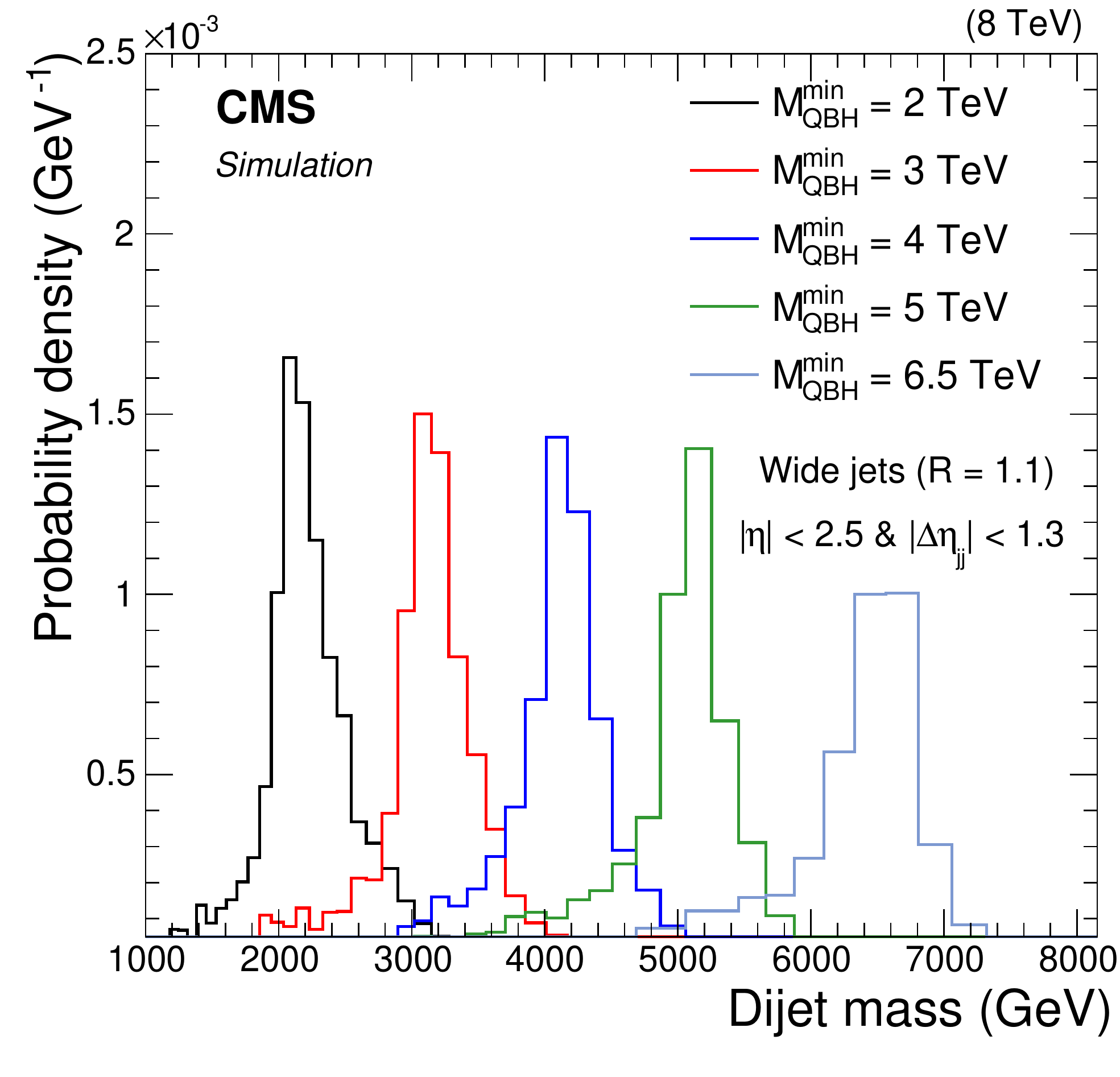}
 \caption{Dijet mass distribution for QBHs with $M_\mathrm{QBH}^\text{min}$ from 2 to 6.5\TeV. The signal shape is almost
independent both of the number of extra dimensions $n$ and the
scale $M_\text{D}$.
\label{fig:BH_mass_spectra}}
\end{figure}

The inclusive dijet search can be interpreted in terms of QBH
production~\cite{MR,Calmet,qbh1} in models with large ($n\ge2$) or
warped ($n=1$) dimensions, where $n$ is the number of extra
dimensions. The dijet invariant mass distribution expected from QBH
decays is used here, in place of the resonance line shape employed in the other analyses.
The required mass shapes are modeled using the \textsc{qbh}
(v1.07) matrix-element generator~\cite{qbh_gen} with the CTEQ6L1
PDF set~\cite{Pumplin:2002vw}, followed by the parton showering
simulation with {\PYTHIA8} and a parametric, fast simulation of the
CMS detector~\cite{FastSim}. The signal is characterized by a peak in
the reconstructed dijet mass spectrum, as shown in
Fig.~\ref{fig:BH_mass_spectra}. The peak position is related to the
minimum mass of QBHs, $M_\mathrm{QBH}^\text{min}$.
The relatively narrow shape is a consequence of the convolution of the threshold-like production behavior for QBHs
with the steeply falling parton luminosity as a function of the
subprocess center-of-mass energy.
The low-mass dijet tails are due
to detector resolution effects.  The signal shape is almost
independent of the number of extra dimensions $n$ and the fundamental
Planck scale $M_\text{D}$. The $n=1$ case corresponds to RS black
holes~\cite{MR}.  In this scenario, $M_\text{D}$ is the product of the
Planck scale and the exponential factor coming from the warping of
the extra dimension.

\begin{figure*}[!htb]
 \centering
   \includegraphics[width=\cmsFigWidth]{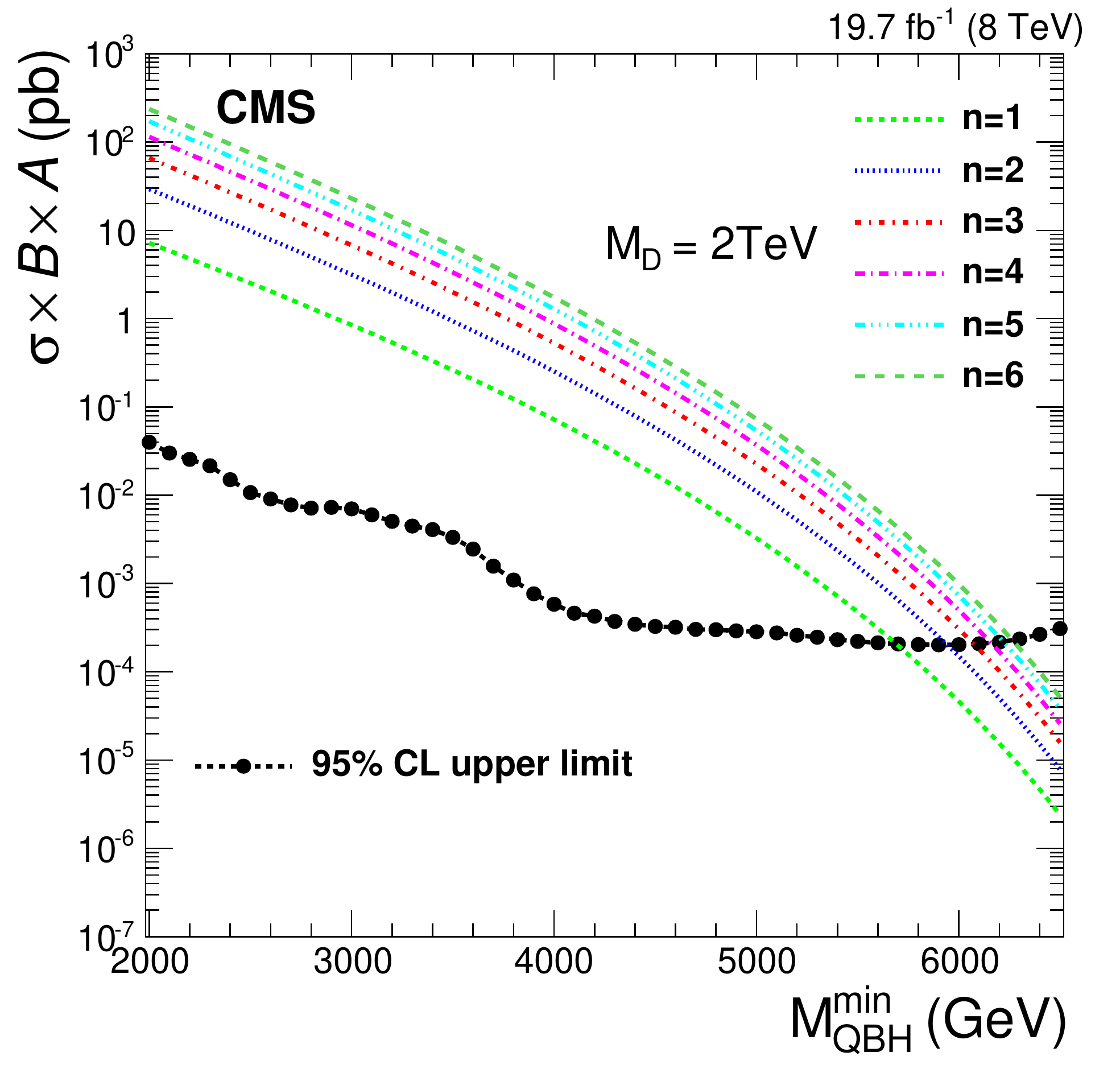}
   \includegraphics[width=\cmsFigWidth]{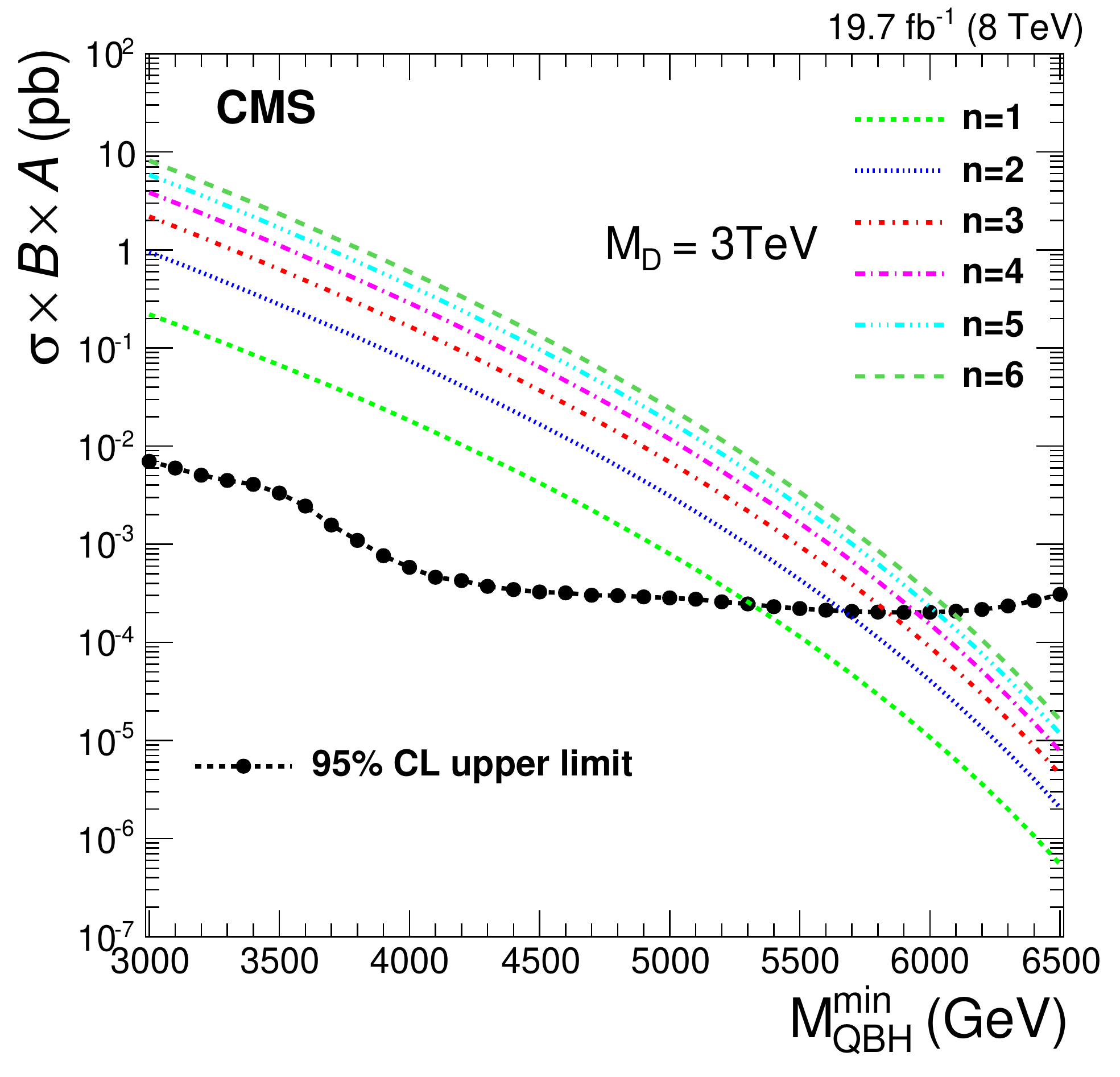}
   \includegraphics[width=\cmsFigWidth]{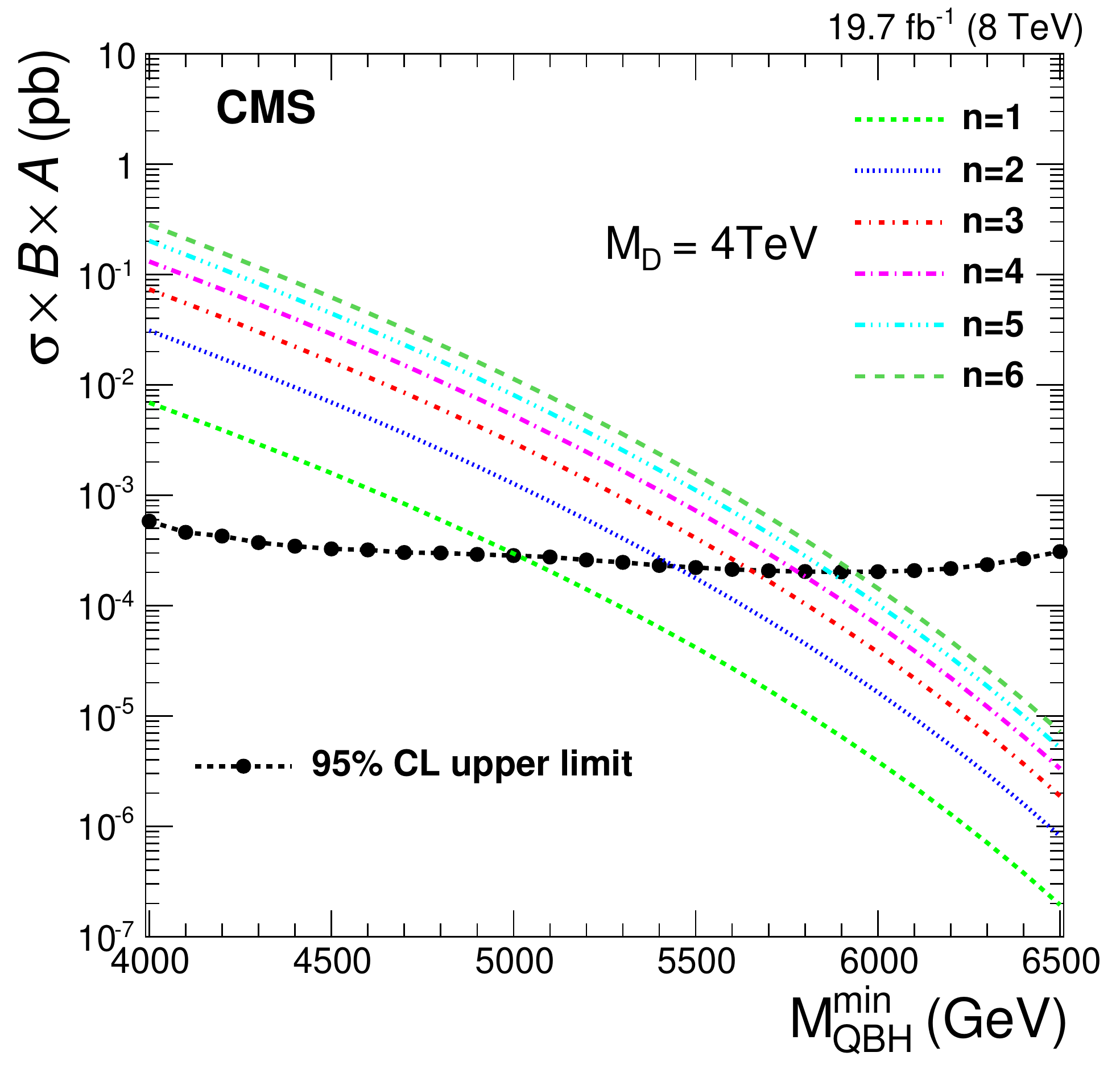}
   \includegraphics[width=\cmsFigWidth]{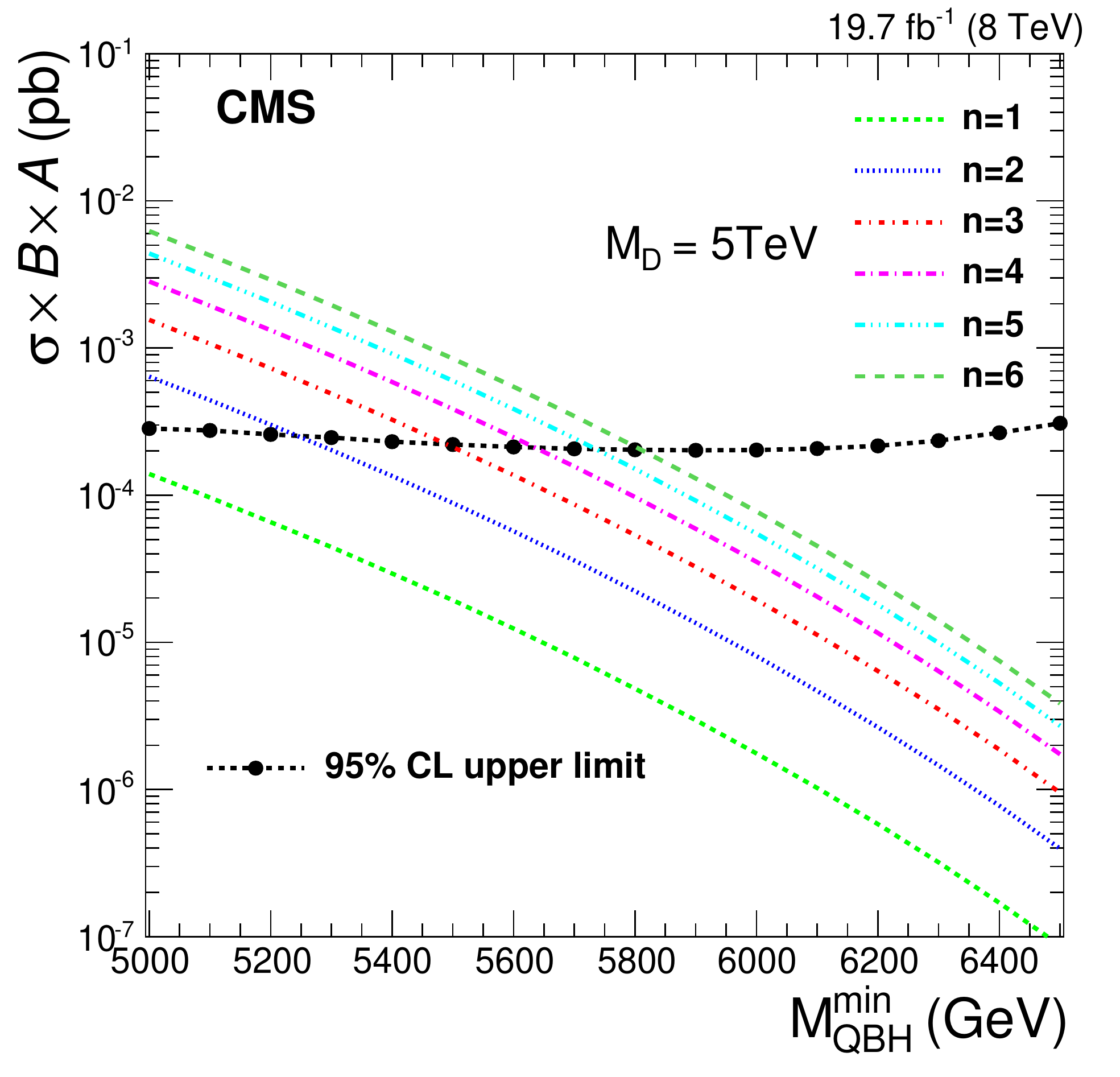}
   \caption{Observed 95\% \CL upper limits on \sba
     as a function of $M_\mathrm{QBH}^\text{min}$, compared to
     theoretical predictions for different values of the fundamental
     Planck scale, $M_\text{D}$, of 2\TeV (top left), 3\TeV (top
     right), 4\TeV (bottom left), and 5\TeV (bottom right), with
     the number of extra dimensions $n$ ranging from 1 to
     6. \label{fig:qbh_limits}}
\end{figure*}

\begin{figure}[!htb]
  \centering
  \includegraphics[width=\cmsFigWidth]{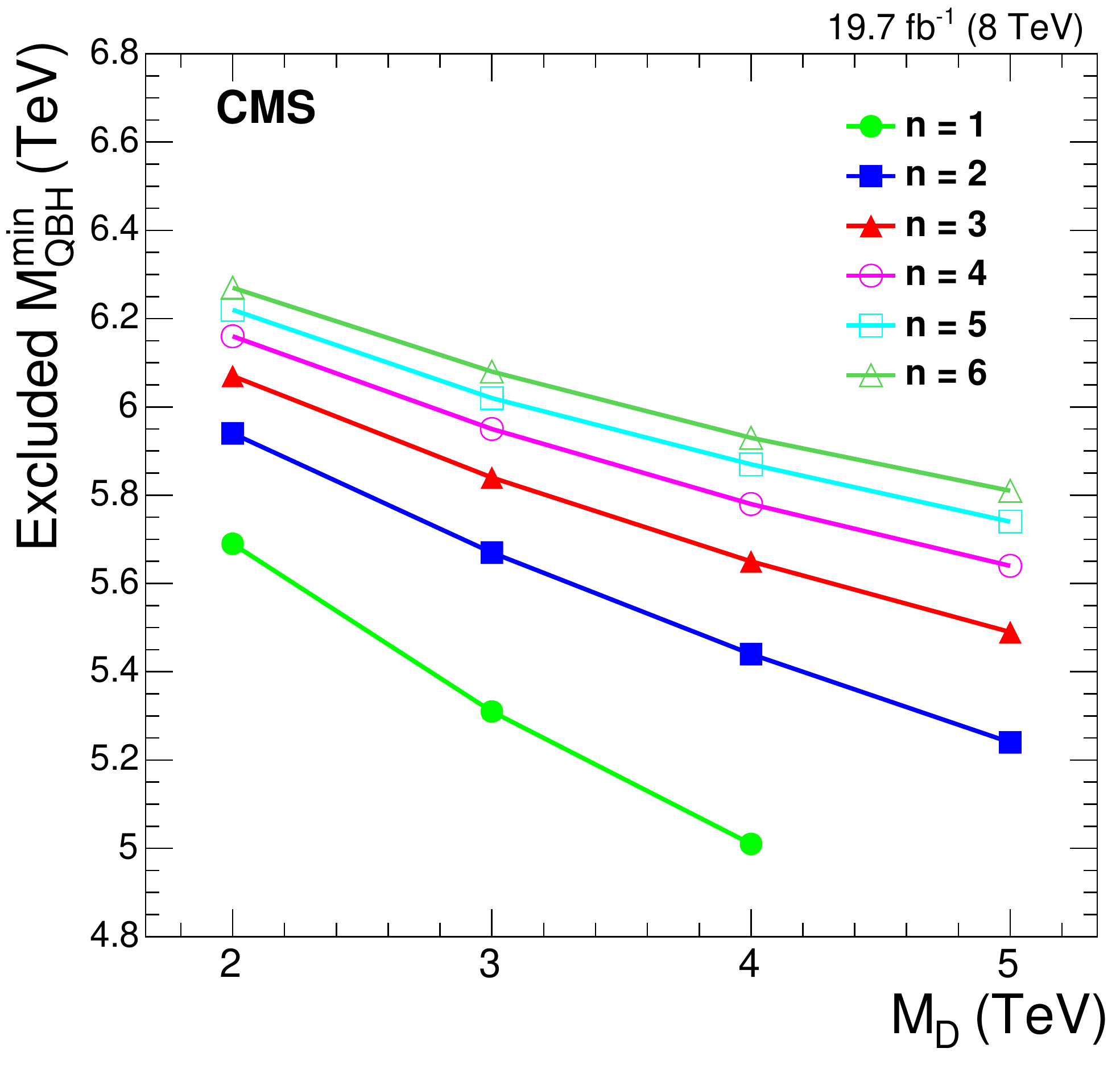}
  \caption{Observed 95\% \CL lower limits on $M_\mathrm{QBH}^\text{min}$ as a function of the Planck
    scale $M_\text{D}$ and the number of extra dimensions $n$.
   \label{fig:qbh_mass}}
\end{figure}

The 95\% \CL observed upper limits on $\sba$ for
QBHs are shown in Fig.~\ref{fig:qbh_limits} and
reported in Table~\ref{tab:limits_obs_inclusive_QBH}. It is commonly
assumed~\cite{Banks:1999gd,Dimopoulos:2001hw,Giddings:2001bu} that
$M_\mathrm{QBH}^\text{min}$ must be greater than or equal to $M_\text{D}$.
Therefore the cross section limits are presented only for
$M_\mathrm{QBH}^\text{min} \ge M_\text{D}$, for different
values of $M_\text{D}$.
The corresponding lower limits on $M_\mathrm{QBH}^\text{min}$
range from 5.0~to~6.3\TeV, depending on the model parameters, and are
shown in Fig.~\ref{fig:qbh_mass} and Table~\ref{tabBH_excludedmass}
as a function of $M_\text{D}$ and $n$. These limits extend those obtained
in Ref.~\cite{QBH2011,Chatrchyan:2013xva}, where the same benchmark
models were considered in the context of a multijet search.

\begin{table*}[!htbp]
  \topcaption{Observed 95\% \CL upper limits on \sba for QBHs from the inclusive
    analysis. These limits are valid for the number of extra dimensions $n$ considered in this paper,
    ranging from 1 to 6. Cross section limits are presented only for
     $M_\mathrm{QBH}^\text{min} \ge M_\text{D}$ for different
values of $M_\text{D}$, as described in the text.\label{tab:limits_obs_inclusive_QBH}}
  \centering
 \begin{scotch}{cccccc}
   $M_\mathrm{QBH}^\text{min}$  &  \multicolumn{5}{c}{Upper limit on \sba\,(fb)} \\
   \multicolumn{1}{c}{(TeV)} &  $M_\text{D}=2$\TeV  &  $M_\text{D}=3$\TeV  &  $M_\text{D}=4$\TeV  &  $M_\text{D}=5$\TeV &  $M_\text{D}=6$\TeV \\ \hline
2 & 40 &	& & &  \\
2.1 & 30 & & & &  \\
2.2 & 25 & & & &  \\
2.3 & 22 & & & &  \\
2.4 & 15 & & & & \\
2.5 & 11 & & & &  \\
2.6 & 9.1 & & & & \\
2.7 & 7.8 & & & &  \\
2.8 & 7.1 & & & &  \\
2.9 & 7.3 & & & &   \\
3 & 7.0 & 7.0 & & &   \\
3.1 & 6.0 & 6.0 &  & &  \\
3.2 & 5.1 & 5.1 &  & &  \\
3.3 & 4.5 & 4.5 &  & &  \\
3.4 & 4.1 & 4.1 & & &  \\
3.5 & 3.3 & 3.3 & & &   \\
3.6 & 2.5 & 2.5 & & &   \\
3.7 & 1.6 & 1.6 & & &   \\
3.8 & 1.1 & 1.1 & & &   \\
3.9 & 0.76 & 0.76 &  & &  \\
4 & 0.58 & 0.58 & 0.58 & &  \\
4.1 & 0.46 & 0.46 & 0.46 & &  \\
4.2 & 0.43 & 0.43 & 0.43 & &   \\
4.3 & 0.37 & 0.37 & 0.37 & &   \\
4.4 & 0.35 & 0.35 & 0.35 & &   \\
4.5 & 0.33 & 0.33 & 0.33 & &   \\
4.6 & 0.32 & 0.32 & 0.32 & &   \\
4.7 & 0.30 & 0.30 & 0.30 & &   \\
4.8 & 0.30 & 0.30 & 0.30 & &   \\
4.9 & 0.29 & 0.29 & 0.29 & &   \\
5 & 0.28 & 0.28 & 0.28 & 0.28 &  \\
5.1 & 0.28 & 0.28 & 0.28 & 0.28 &   \\
5.2 & 0.26 & 0.26 & 0.26 & 0.26 &   \\
5.3 & 0.25 & 0.25 & 0.25 & 0.25 &   \\
5.4 & 0.23 & 0.23 & 0.23 & 0.23 &   \\
5.5 & 0.22 & 0.22 & 0.22 & 0.22 &   \\
5.6 & 0.21 & 0.21 & 0.21 & 0.21 &   \\
5.7 & 0.21 & 0.21 & 0.21 & 0.21 &   \\
5.8 & 0.20 & 0.20 & 0.20 & 0.20 &   \\
5.9 & 0.20 & 0.20 & 0.20 & 0.20 &   \\
6. & 0.20 & 0.20 & 0.20 & 0.20 & 0.20   \\
6.1 & 0.21 & 0.21 & 0.21 & 0.21 & 0.21   \\
6.2 & 0.22 & 0.22 & 0.22 & 0.22 & 0.22  \\
6.3 & 0.23 & 0.23 & 0.23 & 0.23 & 0.23  \\
6.4 & 0.27 & 0.27 & 0.27 & 0.27 & 0.27   \\
6.5 & 0.31 & 0.31 & 0.31 & 0.31 & 0.31  \\
 \end{scotch}
\end{table*}

\begin{table}[htb]
  \topcaption{Observed 95\% \CL lower limits on $M_\mathrm{QBH}^\text{min}$
    for different numbers of extra dimensions
    $n$ and several values of $M_\text{D}$.\label{tabBH_excludedmass}}
\centering
\begin{scotch}{ccccc}
$n$ & \multicolumn{4}{c}{$M_\text{D}$ (\TeVns)} \\
\cline{2-5}
  &  \multicolumn{1}{c}{2}  &  \multicolumn{1}{c}{3} &  \multicolumn{1}{c}{4}  &  \multicolumn{1}{c}{5} \\
\hline
1  & 5.7 & 5.3 & 5.0 & \\
2 & 5.9 & 5.7 & 5.4 & 5.2  \\
3 & 6.1 & 5.8 & 5.7 & 5.5   \\
4 & 6.2 & 6.0 & 5.8 & 5.6   \\
5 & 6.2 & 6.0 & 5.9 & 5.7   \\
6 & 6.3 & 6.0 & 5.9 & 5.8   \\
\end{scotch}
\end{table}

\section{Summary}
A search for resonances and quantum black holes has been performed in inclusive and \PQb-tagged
dijet mass spectra measured with the CMS detector at the LHC. The
data set corresponds to 19.7\fbinv of integrated luminosity
collected in proton-proton collisions at $\sqrt{s}=8$\TeV. The inclusive
search has been conducted for narrow resonances and, for the
first time, for wide resonances with relative widths up to 30\% of the
resonance mass. No evidence for new particle production is found. Upper limits at 95\% \CL
on the product of the cross section, branching fraction into dijets, and
acceptance are provided for all generic searches.  Specific lower
limits are set on the masses of string resonances, excited quarks,
axigluons, colorons, color-octet scalar resonances, scalar diquarks,
\PWpr and $\cPZpr$ bosons, and RS gravitons. For the first time, an exclusion
  limit is set for excited \PQb quarks. The lower mass limits reach
up to 5\TeV, depending on the model, and extend previous exclusions
based on the dijet mass search technique. Quantum black holes up to a mass ranging from
5.0 to 6.3\TeV are also excluded at 95\% \CL, depending on the model.

\begin{acknowledgments}
\hyphenation{Bundes-ministerium Forschungs-gemeinschaft Forschungs-zentren} We congratulate our colleagues in the CERN accelerator departments for the excellent performance of the LHC and thank the technical and administrative staffs at CERN and at other CMS institutes for their contributions to the success of the CMS effort. In addition, we gratefully acknowledge the computing centers and personnel of the Worldwide LHC Computing Grid for delivering so effectively the computing infrastructure essential to our analyses. Finally, we acknowledge the enduring support for the construction and operation of the LHC and the CMS detector provided by the following funding agencies: the Austrian Federal Ministry of Science, Research and Economy and the Austrian Science Fund; the Belgian Fonds de la Recherche Scientifique, and Fonds voor Wetenschappelijk Onderzoek; the Brazilian Funding Agencies (CNPq, CAPES, FAPERJ, and FAPESP); the Bulgarian Ministry of Education and Science; CERN; the Chinese Academy of Sciences, Ministry of Science and Technology, and National Natural Science Foundation of China; the Colombian Funding Agency (COLCIENCIAS); the Croatian Ministry of Science, Education and Sport, and the Croatian Science Foundation; the Research Promotion Foundation, Cyprus; the Ministry of Education and Research, Estonian Research Council via IUT23-4 and IUT23-6 and European Regional Development Fund, Estonia; the Academy of Finland, Finnish Ministry of Education and Culture, and Helsinki Institute of Physics; the Institut National de Physique Nucl\'eaire et de Physique des Particules~/~CNRS, and Commissariat \`a l'\'Energie Atomique et aux \'Energies Alternatives~/~CEA, France; the Bundesministerium f\"ur Bildung und Forschung, Deutsche Forschungsgemeinschaft, and Helmholtz-Gemeinschaft Deutscher Forschungszentren, Germany; the General Secretariat for Research and Technology, Greece; the National Scientific Research Foundation, and National Innovation Office, Hungary; the Department of Atomic Energy and the Department of Science and Technology, India; the Institute for Studies in Theoretical Physics and Mathematics, Iran; the Science Foundation, Ireland; the Istituto Nazionale di Fisica Nucleare, Italy; the Ministry of Science, ICT and Future Planning, and National Research Foundation (NRF), Republic of Korea; the Lithuanian Academy of Sciences; the Ministry of Education, and University of Malaya (Malaysia); the Mexican Funding Agencies (CINVESTAV, CONACYT, SEP, and UASLP-FAI); the Ministry of Business, Innovation and Employment, New Zealand; the Pakistan Atomic Energy Commission; the Ministry of Science and Higher Education and the National Science Centre, Poland; the Funda\c{c}\~ao para a Ci\^encia e a Tecnologia, Portugal; JINR, Dubna; the Ministry of Education and Science of the Russian Federation, the Federal Agency of Atomic Energy of the Russian Federation, Russian Academy of Sciences, and the Russian Foundation for Basic Research; the Ministry of Education, Science and Technological Development of Serbia; the Secretar\'{\i}a de Estado de Investigaci\'on, Desarrollo e Innovaci\'on and Programa Consolider-Ingenio 2010, Spain; the Swiss Funding Agencies (ETH Board, ETH Zurich, PSI, SNF, UniZH, Canton Zurich, and SER); the Ministry of Science and Technology, Taipei; the Thailand Center of Excellence in Physics, the Institute for the Promotion of Teaching Science and Technology of Thailand, Special Task Force for Activating Research and the National Science and Technology Development Agency of Thailand; the Scientific and Technical Research Council of Turkey, and Turkish Atomic Energy Authority; the National Academy of Sciences of Ukraine, and State Fund for Fundamental Researches, Ukraine; the Science and Technology Facilities Council, UK; the US Department of Energy, and the US National Science Foundation.

Individuals have received support from the Marie-Curie program and the European Research Council and EPLANET (European Union); the Leventis Foundation; the A. P. Sloan Foundation; the Alexander von Humboldt Foundation; the Belgian Federal Science Policy Office; the Fonds pour la Formation \`a la Recherche dans l'Industrie et dans l'Agriculture (FRIA-Belgium); the Agentschap voor Innovatie door Wetenschap en Technologie (IWT-Belgium); the Ministry of Education, Youth and Sports (MEYS) of the Czech Republic; the Council of Science and Industrial Research, India; the HOMING PLUS program of Foundation for Polish Science, cofinanced from European Union, Regional Development Fund; the Compagnia di San Paolo (Torino); the Consorzio per la Fisica (Trieste); MIUR project 20108T4XTM (Italy); the Thalis and Aristeia programs cofinanced by EU-ESF and the Greek NSRF; and the National Priorities Research Program by Qatar National Research Fund.
\end{acknowledgments}

\appendix

\section{Cross section calculation for wide resonances}\label{app:xsectionWideRes}

Cross sections for narrow resonances are often given in the narrow width
approximation, where the sub-process cross section
\begin{equation}
\hat{\sigma}(\hat{s}) \propto \delta (\hat{s} - M_X^2)
\label{delta-function}
\end{equation}
is integrated over the PDFs (Section 2.2.11 in
Ref.~\cite{Harris:2011bh}).  Here $\hat{s}=m^2$ is the square of the
diparton mass, $M_X$ is the resonance mass, and the delta function
implies that the PDFs are evaluated at only those values of fractional
momenta $x_1$ and $x_2$ that correspond to the resonance pole:
$M_X^2=\hat{s}=x_1 x_2 s$, where $s$ is the square of the
proton-proton collision energy. Cross sections calculated in the
narrow-width approximation are appropriate for comparison to CMS upper
limits for narrow resonances, because the dijet mass resonance shapes used in that search
correspond to a relative resonance width ($\Gamma/M$) much
smaller than the detector resolution.

Multiple processes can contribute to the total cross section for wide resonance
production. The $s$-channel process, the annihilation of two initial state partons
into the resonance, is usually the most significant contribution and is the process
searched for by this analysis. The $s$-channel cross section is evaluated by replacing
the delta function in Eq.~(\ref{delta-function}) with a full
relativistic Breit--Wigner resonance shape,
before integrating over the PDFs. The $t$-channel process,
where the new particle is exchanged between the incoming partons, often has an appreciable
contribution to the cross section but it does not peak sharply in diparton mass and
may be absorbed into the background shape during a search. The interference process,
including interference between the multijet background processes and both the $s$- and
$t$-channel signal processes, could often significantly modify the resonance shape far
off the resonance pole. Interference contributions depend on the type
of resonance considered, and are not included in the resonance shape used in our search.
Our calculation of the wide resonance cross section
is an approximation that considers only the $s$-channel term, to
which limits from our search should be compared.

The cross section calculations for wide resonances employ a resonance
shape for the $s$-channel resonances as a function of $\hat{s} -
M_X^2$.  In order for this calculated cross section to be comparable
to the resonance upper limits, we have used the same shape for the
underlying parton-parton scattering sub-process cross section as is
used in the search to set limits. The shape corresponds to an RS
graviton resonance.  The generator used, {\sc pythia6}, models that
shape with the following, general Breit--Wigner resonance formula (Eq.~(7.47)
in~\cite{Sjostrand:2006za}):
\begin{equation}
\hat{\sigma}_{i \to R \to f}(\hat{s}) \propto \frac{\pi}{\hat{s}} \,
\frac{H_R^{(i)}(\hat{s}) \, H_R^{(f)}(\hat{s})}
{(\hat{s} - M_X^2)^2 + H_R^2(\hat{s})},
\label{eq:Breit-Wigner}
\end{equation}
where
\begin{equation}
H_R(\hat{s}) = \frac{\hat{s} \Gamma_R}{M_X}
\label{eq:Htotal}
\end{equation}
and $\Gamma_R$ is the full resonance width.
For the RS graviton resonance
\begin{equation}
H_R^{(i,f)}(\hat{s}) = \left ( \frac{\hat{s}}{M_X^2} \right ) \frac{\hat{s} \Gamma_R^{(i,f)}}{ M_X}
\label{eq:Hif}
\end{equation}
where $\Gamma_R^{(i,f)}$ are the partial widths for the initial state
$i$ and final state $f$.  We note that the term $\hat{s}/M_X^2$ in
Eq.~(\ref{eq:Hif}) significantly affects the resonance shape
far away from the resonance pole, suppressing the tail at low diparton mass.
This term is appropriate for resonances that have a width proportional to the
cube of the resonance mass, like the RS graviton or the color-octet
scalar. Even with this suppression, the enhancement at low dijet mass due to convolution
of the tail with PDFs is visible in Fig.~\ref{fig:shapes_broadResonance}
for resonances with the highest widths and masses.

We calculate the full wide resonance cross section from $s$-channel
production by integrating the Breit--Wigner resonance shape defined by
Eqs.~(\ref{eq:Breit-Wigner})--(\ref{eq:Hif}), over the PDFs.
Table~\ref{tab_gg_xsec} shows the full cross section divided by the
cross section in the narrow-width approximation as a function of the
resonance mass and width, for both $\PQq\PAQq$ and $\Pg\Pg$ resonances.  This ratio is close to unity for narrow resonances, for
which the full cross section and the narrow-width approximation cross
section are naturally the same. For wide resonances at high resonance
mass this ratio can be significantly greater than 1, because the
convolution of the PDFs with the low mass tail of the Breit--Wigner result
in a large cross section at low diparton mass.  For wide resonances
the values reported in Table~\ref{tab_gg_xsec} can be applied as a
multiplicative correction to the narrow-width approximation cross
sections to get an appropriate resonance cross section to compare with
our experimental upper limits on cross section.  This is done in
Fig.~\ref{fig:limit_axigluon_S8} to obtain the model cross section
presented and to set mass limits for axigluons/colorons and
color-octet scalars.  The correction factor for axigluons, which are
$\PQq\PAQq$ resonances of width $\Gamma_R=\alpha_S M_X$,
is 1.1 at a mass of 3.5\TeV.  The correction factor for color-octet
scalars, which are $\Pg\Pg$ resonances of width $\Gamma_R=5\alpha_S M_X/6$,
is 1.1 at a mass of 2.5\TeV. So for these resonances, at mass values
close to our mass limit, the full cross section is close to the cross
section calculated in the narrow-width approximation. We recommend the
same procedure, using Table~\ref{tab_gg_xsec}, for users of our limits
on the wide-resonance cross section, as this will ensure that the
resonance shape used to calculate the cross section matches the
resonance shape we used to set limits.

For resonances with widths that are directly proportional
to the resonance mass, like axigluons or colorons, the following term
is normally used instead of Eq.~(\ref{eq:Hif}) to describe the resonance line shape:
\begin{equation}
H_R^{(i,f)}(\hat{s}) =  \frac{\hat{s} \Gamma_R^{(i,f)}}{M_X}.
\label{eq:VectorPartialWidth}
\end{equation}
For many wide resonances of interest this term produces a resonance shape with a
very large tail at low mass: a cross section that falls rapidly
with increasing diparton mass, like the multijet background.  This
shape at low dijet mass would be largely absorbed into the multijet background definition
of our search.  Like the multijet background, the full cross section for this
wide shape is mainly determined by the lowest diparton mass considered.
This shape is therefore difficult to use in a well
defined fashion in a search that sets upper limits on a resonance cross section,
because the cross section is only weakly dependent on the resonance pole mass. Thus,
we have limited the wide resonance search to the shape defined
by Eq.~(\ref{eq:Hif}). Our wide resonance search results are still applicable
for a range of resonance widths and masses even for resonances that have a shape
defined by Eq.~(\ref{eq:VectorPartialWidth}).  As long as the full cross section
for the true resonance line shape integrated over the mass interval of the
CMS search is not larger than about 20 times the narrow-width approximation cross section,
the results of the CMS search are approximately valid and applicable.
This approximate range of validity is derived by comparing
Table~\ref{tabggfastLimit_newrange} with~\ref{tab_gg_xsec}.
The boundary of validity of the limits shown in Table~\ref{tabggfastLimit_newrange}
has an average ratio value of about 20 in Table~\ref{tab_gg_xsec}.
Note that our limits are valid if the condition reported in Section~\ref{sec:WideRes}
holds so that the low mass tail does not significantly affect the shape analysis.
Thus, to first approximation, only this comparison of the full
resonance cross section to the narrow width cross section matters in determining
validity of our limits, independent of the shape of the low mass tail.

\begin{table}[htb]
  \topcaption{
    Correction factors defined as the ratio of the full cross section obtained from
    Eqs.~(\ref{eq:Breit-Wigner})-(\ref{eq:Hif}) to the cross section
    from the narrow-width approximation calculations, as a function of the
    resonance mass, for \PQq\PAQq and $\Pg\Pg$ resonances and for eight different
    resonance widths in proton-proton collisions at $\sqrt{s}=8$\TeV.\label{tab_gg_xsec}}
\centering
\begin{scotch}{ccccccccc}
Mass   &  \multicolumn{8}{c}{Fractional width ($\Gamma$/M)}\\
\cline{2-9}
 (\TeVns)  & $10^{-5}$ & 0.015 & 0.05  & 0.1   & 0.15  & 0.2   & 0.25  & 0.3   \\ \hline
\multicolumn{9}{c}{Correction factors for \PQq\PAQq resonances}\\ \hline
  1.25 &   1.00 &   1.00 &   0.97 &   0.85 &   0.81 &   0.79 &   0.77 &   0.76 \\
  1.50 &   1.00 &   1.00 &   0.97 &   0.85 &   0.81 &   0.79 &   0.77 &   0.77 \\
  1.75 &   1.00 &   1.00 &   0.97 &   0.85 &   0.82 &   0.79 &   0.78 &   0.78 \\
  2.00 &   1.00 &   1.01 &   0.98 &   0.86 &   0.83 &   0.81 &   0.80 &   0.80 \\
  2.25 &   1.00 &   1.01 &   0.98 &   0.87 &   0.84 &   0.83 &   0.83 &   0.84 \\
  2.50 &   1.00 &   1.01 &   0.99 &   0.89 &   0.87 &   0.87 &   0.88 &   0.89 \\
  2.75 &   1.00 &   1.01 &   1.01 &   0.92 &   0.91 &   0.92 &   0.94 &   0.97 \\
  3.00 &   1.00 &   1.01 &   1.03 &   0.96 &   0.97 &   1.00 &   1.04 &   1.09 \\
  3.25 &   1.00 &   1.02 &   1.05 &   1.02 &   1.06 &   1.11 &   1.18 &   1.26 \\
  3.50 &   1.00 &   1.02 &   1.09 &   1.12 &   1.20 &   1.29 &   1.40 &   1.52 \\
  3.75 &   1.00 &   1.03 &   1.14 &   1.26 &   1.41 &   1.57 &   1.74 &   1.92 \\
  4.00 &   1.00 &   1.04 &   1.22 &   1.50 &   1.75 &   2.02 &   2.30 &   2.60 \\
  4.25 &   1.00 &   1.05 &   1.34 &   1.91 &   2.35 &   2.81 &   3.29 &   3.80 \\
  4.50 &   1.00 &   1.06 &   1.54 &   2.67 &   3.46 &   4.27 &   5.12 &   6.03 \\
  4.75 &   1.00 &   1.08 &   1.87 &   4.16 &   5.66 &   7.20 &   8.82 &  10.5 \\
  5.00 &   1.00 &   1.11 &   2.45 &   7.36 &  10.4  &  13.5  &   16.9 &  20.4 \\
  5.25 &   1.00 &   1.14 &   3.52 &  14.8  &  21.4  &  28.4  &   35.7 &  43.6 \\
  5.50 &   1.00 &   1.19 &   5.60 &  33.3  &  49.1  &  65.6  &   83.2 & 102 \\ \hline
\multicolumn{9}{c}{Correction factors for $\Pg\Pg$ resonances}\\ \hline
  1.25 &   1.00 &   1.01 &   0.98 &   0.88 &   0.85 &   0.84 &   0.84 &   0.85 \\
  1.50 &   1.00 &   1.01 &   0.99 &   0.89 &   0.87 &   0.87 &   0.88 &   0.90 \\
  1.75 &   1.00 &   1.01 &   1.00 &   0.92 &   0.91 &   0.92 &   0.94 &   0.96 \\
  2.00 &   1.00 &   1.01 &   1.01 &   0.95 &   0.95 &   0.98 &   1.01 &   1.06 \\
  2.25 &   1.00 &   1.01 &   1.03 &   0.99 &   1.02 &   1.06 &   1.12 &   1.18 \\
  2.50 &   1.00 &   1.02 &   1.05 &   1.06 &   1.11 &   1.18 &   1.27 &   1.37 \\
  2.75 &   1.00 &   1.02 &   1.08 &   1.15 &   1.24 &   1.36 &   1.49 &   1.63 \\
  3.00 &   1.00 &   1.02 &   1.11 &   1.28 &   1.44 &   1.62 &   1.81 &   2.03 \\
  3.25 &   1.00 &   1.03 &   1.16 &   1.49 &   1.74 &   2.02 &   2.31 &   2.64 \\
  3.50 &   1.00 &   1.03 &   1.23 &   1.81 &   2.22 &   2.65 &   3.11 &   3.61 \\
  3.75 &   1.00 &   1.04 &   1.33 &   2.34 &   3.00 &   3.70 &   4.44 &   5.23 \\
  4.00 &   1.00 &   1.05 &   1.46 &   3.26 &   4.36 &   5.52 &   6.75 &   8.07 \\
  4.25 &   1.00 &   1.06 &   1.67 &   4.94 &   6.86 &   8.87 &  11.01 &  13.3 \\
  4.50 &   1.00 &   1.08 &   1.99 &   8.20 &  11.7  &  15.4  &  19.4  &  23.6 \\
  4.75 &   1.00 &   1.10 &   2.51 &  15.0  &  21.9  &  29.1  &  36.9  &  45.3 \\
  5.00 &   1.00 &   1.12 &   3.43 &  30.4  &  44.9  &  60.2  &  76.6  &  94.5 \\
  5.25 &   1.00 &   1.16 &   5.17 &  68.3  & 102    & 137    & 175    & 216 \\
  5.50 &   1.00 &   1.22 &   8.73 & 172    & 257    & 348    & 445    & 552 \\
\end{scotch}
\end{table}

\bibliography{auto_generated}   

\cleardoublepage \section{The CMS Collaboration \label{app:collab}}\begin{sloppypar}\hyphenpenalty=5000\widowpenalty=500\clubpenalty=5000\textbf{Yerevan Physics Institute,  Yerevan,  Armenia}\\*[0pt]
V.~Khachatryan, A.M.~Sirunyan, A.~Tumasyan
\vskip\cmsinstskip
\textbf{Institut f\"{u}r Hochenergiephysik der OeAW,  Wien,  Austria}\\*[0pt]
W.~Adam, T.~Bergauer, M.~Dragicevic, J.~Er\"{o}, M.~Friedl, R.~Fr\"{u}hwirth\cmsAuthorMark{1}, V.M.~Ghete, C.~Hartl, N.~H\"{o}rmann, J.~Hrubec, M.~Jeitler\cmsAuthorMark{1}, W.~Kiesenhofer, V.~Kn\"{u}nz, M.~Krammer\cmsAuthorMark{1}, I.~Kr\"{a}tschmer, D.~Liko, I.~Mikulec, D.~Rabady\cmsAuthorMark{2}, B.~Rahbaran, H.~Rohringer, R.~Sch\"{o}fbeck, J.~Strauss, W.~Treberer-Treberspurg, W.~Waltenberger, C.-E.~Wulz\cmsAuthorMark{1}
% ...
\vskip\cmsinstskip
% ...
1:~~Also at Vienna University of Technology, Vienna, Austria\\
2:~~Also at CERN, European Organization for Nuclear Research, Geneva, Switzerland\\
% ...

\end{sloppypar}
\end{document}